\documentclass[aps,prb,reprint,groupedaddress, showkeys]{revtex4-2}
\usepackage{physics}
\usepackage{siunitx}
\usepackage{mathtools}
\usepackage{amssymb}
\usepackage{bm}      % Bold math
\usepackage{graphicx}
\graphicspath{{./Figures/}}
\usepackage{hyperref}
\hypersetup{unicode=true, colorlinks=true, linkcolor=blue, citecolor=blue}
\usepackage{titlesec}
\titlespacing*{\subsection}{0pt}{*3}{0pt} % {left}{before}{after}

\usepackage{enumitem}
\usepackage{textcase}
\makeatletter

\def\rtx@sectionbase{%
  \@startsection{section}{1}{\z@}%
    {1\baselineskip}% space before
    {1\baselineskip}% space after
    {\normalfont\bfseries\centering}% <- center the heading
}

\renewcommand{\section}{\@ifstar{\rtx@sectstar}{\rtx@sect}}
\newcommand{\rtx@sect}[2][]{%
  \rtx@sectionbase[\MakeTextUppercase{#1}]{\MakeTextUppercase{#2}}%
}
\newcommand{\rtx@sectstar}[1]{%
  \rtx@sectionbase*{\MakeTextUppercase{#1}}%
}

\makeatother

\begin{document}
\preprint{APS/123-QED}

\title{Entropy Production from Spin--Vibrational Coupling in Endohedral-Fullerene Qubits Encapsulated in Suspended Carbon Nanotubes}
\author{Cristian Staii}%
\email{cstaii01@tufts.edu}
\affiliation{Department of Physics and Astronomy, Tufts University, Medford, MA, 02155, USA}

\date{\today}

\begin{abstract}
Hybrid carbon nanotube--fullerene architectures provide a controllable setting in which to study irreversibility and information flow in strongly structured quantum environments. We analyze entropy generation in a platform where paramagnetic endohedral fullerenes (PEFs), such as N@C$_{60}$ and P@C$_{60}$, are encapsulated inside a suspended carbon nanotube (CNT) resonator, such that selected multi-level PEF spin states define an effective qubit coupled to quantized CNT flexural modes. Motivated by prior work on fullerene-filled CNTs, on spin--phonon manipulation in suspended nanotubes, and on exact phase-space propagators for damped driven oscillators, we formulate a hybrid open-system description that combines a driven quantum Brownian description of the CNT resonator with an effective Jaynes--Cummings type spin--vibrational interaction. The resonator dynamics are represented in phase space through the Wigner function, whose time evolution can be written analytically in terms of the initial Wigner distribution and a Gaussian propagator. This representation makes it possible to separate drive-induced phase space displacement, diffusion, and damping, and to connect these features directly to entropy flow. The coupled spin--mechanical dynamics are then embedded in a Lindblad quantum master equation that includes mechanical damping, spin relaxation, pure dephasing, and thermally activated excitation channels. Within this framework we derive the entropy balance equation---identifying entropy flux and non-negative entropy production---and examine how hybridization between the molecular spin and the nanotube vibration redistributes irreversibility between coherent exchange and dissipative channels. We show that spin--phonon coupling enhanced by a magnetic field gradient, resonant driving, and moderate thermal occupation can produce identifiable crossovers between entropy--production regimes dominated by the oscillator and those dominated by the spin. The resulting framework provides a quantitative basis for using CNT--PEF hybrids as nanoscale platforms for studying nonequilibrium quantum thermodynamics, decoherence, and information loss in structured vibrational~environments.
\end{abstract}
% Keywords
\keywords{endohedral fullerenes; suspended carbon nanotubes; spin--phonon coupling; Wigner function; quantum Brownian motion; entropy production; irreversibility; Lindblad dynamics; open quantum systems; quantum thermodynamics}%Use showkeys class option if keyword display desired
\maketitle

\section{Introduction}

The controlled generation, storage, and~degradation of quantum information in hybrid nanostructures lie at the center of both quantum technologies and nonequilibrium statistical physics~\cite{Landi:2020bsq, Zurek:2003zz, schlosshauer_decoherence_2007, lee_strong_2023, samanta_nonlinear_2023, wollack_quantum_2022, rossi_measurement-based_2018, blencowe_quantum_2004, lahaye_approaching_2004}. Of~particular interest are systems in which localized spin degrees of freedom interact with vibrational modes that are both strongly confined and experimentally accessible. In~such systems, irreversibility is not merely a secondary effect but also a source of information about the pathways through which energy, coherence, and~entropy are redistributed among subsystems and reservoirs~\cite{palyi_spinorbitinduced_2012, wang_creating_2016, wang_method_2016, boissonneault_dispersive_2009, cui_feedback_2013, riste_initialization_2012, aspelmeyer_cavity_2014}.

Endohedral fullerenes, especially paramagnetic endohedral fullerenes (PEFs) such as N@C$_{60}$ and P@C$_{60}$, are especially attractive molecular spin units because they combine long spin coherence times with chemically robust carbon cages and a rich internal level structure~\cite{pietzak_endofullerenes_2002, porfyrakis_endohedral_2016, morton_relaxation_2006}. Their encapsulation inside carbon nanotubes leads to ``peapod'' architectures in which the fullerenes are spatially confined, electronically addressable, and~mechanically embedded in a quasi-one-dimensional (1D) host~\cite{simon_low_2004, eckardt_stability_2015, waiblinger_thermal_2001, morton_environmental_2007}. Such fullerene-filled nanotubes have been investigated as candidate architectures for quantum information processing, spin transport, and~engineered exchange interactions~\cite{harneit_fullerenes_2017, rips_hartmann_prl_2013, pinto_readout_2020}. In~this setting, the~nanotube is not merely a passive scaffold: when suspended, it supports quantized flexural modes that can couple to internal spins either intrinsically, via spin--orbit and deflection-mediated mechanisms, or extrinsically, or~through externally applied magnetic-field gradients~\cite{palyi_spinorbitinduced_2012, wang_creating_2016}.

Suspended CNT resonators provide a particularly compelling platform for studying spin--vibrational irreversibility because they naturally combine three features. First, their flexural modes can be coherently driven and detected with high sensitivity~\cite{sazonova_tunable_2004, garcia-sanchez_mechanical_2007, huttel_nanoelectromechanics_2008, staii_high_frequency_sgm_2005, staii_dna_decorated_2005, eichler_nonlinear_2011, tavernarakis_optomechanics_2018}. Second, their coupling to localized states can be tuned into regimes described by effective Jaynes--Cummings or Rabi-type models~\cite{schneider_observation_2014, wang_method_2016, sarma_tunable_2018, wang_hybrid_2017}. Third, their interaction with phononic, electronic, and~electromagnetic environments can be captured within open-system formalisms that expose entropy flow at the quantum level~\cite{semiao_kerr_2009, rips_nonlinear_2014, Steele2009Science, willick_probing_2017, Urgell2020NatPhys}. Previous work on suspended CNT spin qubits has also shown that mechanical motion can mediate electrically controllable spin resonance and coherent spin manipulation~\cite{palyi_spinorbitinduced_2012, wang_creating_2016}. At~the same time, exact phase space methods developed for the damped driven quantum oscillator provide a transparent way to treat the mechanical mode in terms of its Wigner function, with~diffusion, damping, and~forcing entering as distinct ingredients of the propagator~\cite{PhysRevD.64.105020, qiu_quantum_2021, breuer_theory_2009}.

In recent work, our group developed a unified all-mechanical protocol for coherent control and quantum-state reconstruction of suspended CNT resonators operated in the anharmonic regime~\cite{chang_quantum_2025}. Using a nearby AFM tip as a single localized actuator, we showed that the fundamental flexural mode can be driven through Rabi and Ramsey sequences, thereby allowing direct extraction of the relaxation and coherence times $T_1$ and $T_2$, while the same actuator can implement controlled phase space displacements for Wigner function tomography via displaced-parity measurements. The~analysis also identified the conditions under which regions with negative Wigner function and other nonclassical signatures remain observable and~provided realistic parameter estimates showing that such measurements are feasible in cryogenic CNT devices with present-day experimental~capabilities. 

The present work builds on these developments to address entropy production in a hybrid system composed of PEF spin qubits encapsulated in a suspended CNT resonator (the PEF-CNT system). Our aim is to construct a physically grounded analytical framework that links four elements: (i) the PEF-CNT architecture and its spin content; (ii) the driven, damped motion of the suspended CNT; (iii) the effective spin--phonon coupling between selected PEF states and a CNT flexural mode; (iv) the entropy balance associated with the resulting Lindblad~dynamics.

The central conceptual point is that the CNT mechanical mode acts simultaneously as a controllable quantum subsystem and as an intermediary through which structured dissipation influences the PEF spin. In~phase space, the~mechanical state evolves by convolution with an exact Gaussian propagator, making visible the separate roles of reversible phase space damping, thermal broadening, and~drive-induced displacement. Once coupled to the spin sector, these processes enter the entropy balance through the reduced density operator of the joint spin--mechanical system. This enables a decomposition of irreversibility into entropy flux toward the reservoirs and intrinsic entropy production associated with departures from detailed~balance.

The paper is organized as follows. Section~\ref{sec:platform} summarizes the physical ingredients of PEF-filled suspended CNTs and motivates the effective qubit description. Section~\ref{sec:model} develops the hybrid Hamiltonian and the open-system master equation. Section~\ref{sec:wigner} reformulates the oscillator sector in terms of the exact Wigner propagator for driven quantum Brownian motion and shows how this solution is embedded into the coupled dynamics. Section~\ref{sec:entropy} derives the entropy balance and identifies entropy flux and entropy production. Section~\ref{sec:results} analyzes representative parameter regimes and discusses crossovers between mechanically and spin-dominated irreversibility. Section~\ref{sec:discussion} discusses the implications for experiments and for nonequilibrium quantum thermodynamics in nanoscale hybrid systems. Section~\ref{sec:Conclusions} presents the main conclusions of this work.

%%%%%%%%%%%%%%%%%%%%%%%%%%%%%%%%%%%%%%%%%%

\section{Physical Platform: Endohedral Fullerenes in Suspended Carbon Nanotubes}
\label{sec:platform}

\begin{figure}
    \includegraphics[width=0.8\linewidth]{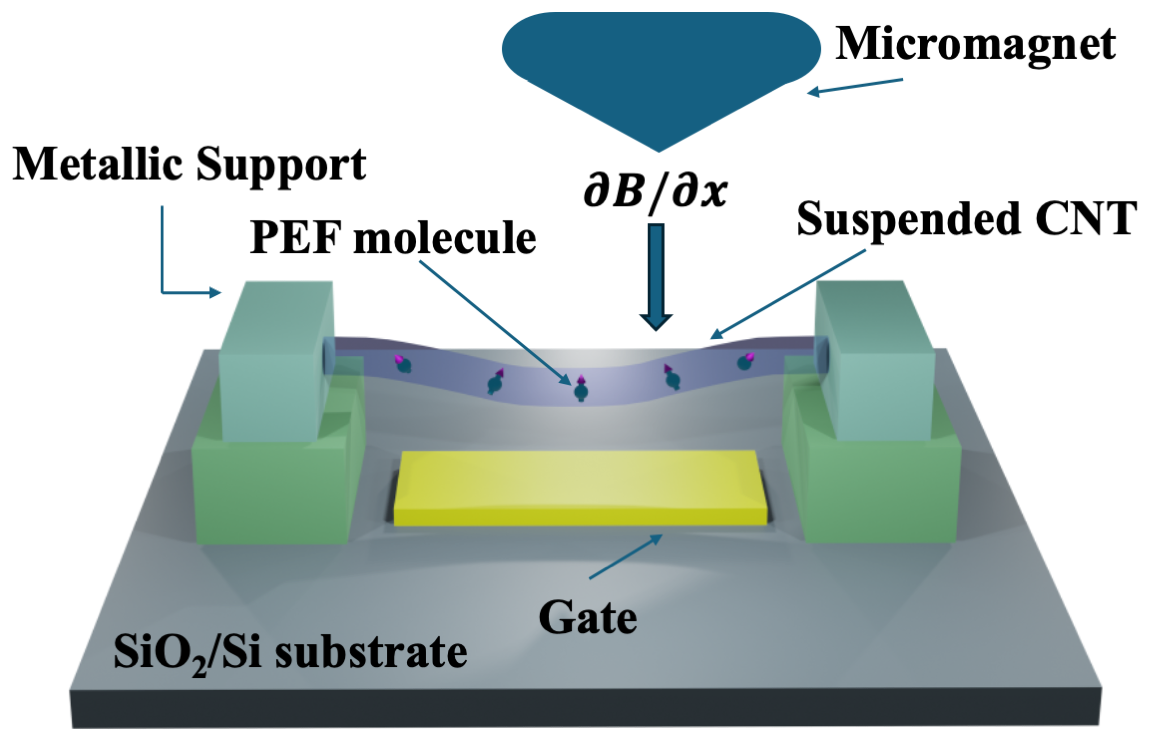}
    \caption{Schematic of the proposed experimental setup showing single spins trapped inside a suspended carbon nanotube (CNT). The spins are associated with paramagnetic endohedral~fullerene (PEF) molecules confined within the CNT. The~vibrational motion of the nanotube is controlled by gate voltages, while a micromagnet placed nearby generates a non-uniform external magnetic field. The~figure indicates the relevant magnetic-field gradient, $\partial B/\partial x$, where $B$ is the component of the micromagnet field along the PEF spin quantization axis and $x$ is the CNT flexural displacement coordinate.  This field gradient makes the PEF spin splitting position dependent, giving rise to spin--vibrational coupling between the PEF spin degree of freedom and the quantized mechanical motion of the nanotube and enabling mechanical control of the PEF spin splitting.   \label{fig1}}
\end{figure}
\unskip

\subsection{PEF-filled carbon nanotubes as hybrid quantum architectures}

Paramagnetic endohedral fullerenes consist of atoms or small clusters trapped inside a fullerene cage~\cite{porfyrakis_endohedral_2016, pietzak_endofullerenes_2002}. In~the families N@C$_{60}$ and P@C$_{60}$, the~enclosed atom retains an unpaired electronic spin that is well shielded by the carbon cage from environmental perturbations, giving rise to narrow spin resonances and long coherence times~\cite{morton_relaxation_2006, simon_low_2004, eckardt_stability_2015, waiblinger_thermal_2001, morton_environmental_2007}. The~use of fullerene cages as molecular spin carriers has motivated extensive proposals for quantum information processing, including arrays of spin-bearing fullerenes assembled within carbon nanotubes~\cite{harneit_fullerenes_2017}.

Suspended CNT resonators provide exceptionally low effective mass, large vibrational motion, and~high mechanical frequencies spanning MHz–GHz, thus enabling ground-state operation and strong coupling to spins~\cite{sazonova_tunable_2004, garcia-sanchez_mechanical_2007, huttel_nanoelectromechanics_2008, staii_high_frequency_sgm_2005, staii_dna_decorated_2005, eichler_nonlinear_2011, tavernarakis_optomechanics_2018,schneider_observation_2014}. When inserted into CNTs, endohedral fullerenes form ordered or quasi-ordered 1D chains in which intermolecular spacing, host--guest interactions, and~nanotube confinement can all influence spin dynamics (Figure~\ref{fig1}). Experimental and theoretical studies of fullerene peapods have highlighted several relevant features: (i) the CNT can stabilize the fullerene chain; (ii) it can modify electronic and magnetic interactions through confinement and charge redistribution; (iii) it can serve as a transport channel or an electromechanical element. For~quantum-control applications, the~fullerene spin can therefore be modeled as interacting with a structured environment that is intermediate between a local environment and a continuum~reservoir.

In this work we focus on a suspended segment of CNT hosting several PEF molecules and~concentrate fullerenes whose low-energy spin manifold defines an effective qubit. The~CNT's lowest flexural mode acts as the principal bosonic mode, while other vibrational modes and external reservoirs are treated as dissipative environments. This reduced description is appropriate when one flexural mode is tuned near resonance with a controllable spin splitting and remains spectrally well separated from the rest of the mechanical~spectrum.

\subsection{Effective Spin Manifold of Endohedral Fullerene~Qubits}
Both N@C$_{60}$ and P@C$_{60}$ exhibit high-spin electronic ground states originating from the encapsulated atom, together with hyperfine coupling to a nuclear spin~\cite{harneit_fullerenes_2017}. For~entropy-production analysis it is sufficient to project this richer manifold onto two states, $\{|g\rangle, |e\rangle\}$, which are selected by external magnetic fields and microwave-frequency addressing. The~resulting {splitting} %MDPI: Please ensure all variables/values in the equation appear in the same format in the text (normal/italic/bold/subscript/superscript).
\begin{equation}
\hbar\omega_s = E_e - E_g,
\end{equation}
can be tuned by a static magnetic field, although~hyperfine structure and anisotropies may generate additional nearby transitions. Our treatment therefore assumes that either those additional levels are far detuned from the mechanical mode and external drive, or~they contribute only perturbatively through renormalized rates and~couplings.

This effective qubit approximation is standard in hybrid spin--oscillator descriptions~\cite{palyi_spinorbitinduced_2012, wang_creating_2016} and is particularly useful for identifying entropy pathways. In~particular, it permits one to distinguish clearly between entropy associated with spin populations and coherences, on~the one hand, and~entropy associated with the vibrational phase-space distribution, on~the other~\cite{breuer_theory_2009}.

\subsection{Suspended CNT resonator and external control}

A suspended CNT behaves as a nanomechanical resonator with flexural frequencies ranging from the MHz to GHz regimes, depending on its length, tension, and~electrostatic environment~\cite{sazonova_tunable_2004, garcia-sanchez_mechanical_2007, huttel_nanoelectromechanics_2008}. In~the harmonic approximation, the~relevant mechanical mode is described by the Hamiltonian~\cite{chang_quantum_2025}
\vspace{-3pt}\begin{equation}
H_m = \hbar\omega_m a^{\dagger}a,
\end{equation}
where \(\omega_m\) is the angular frequency of the CNT flexural mode, and~ $a$ and $a^{\dagger}$ are bosonic annihilation and creation operators. Coherent driving can be implemented electrically or magnetically by applying, for~example, an ac gate voltage that produces a time-dependent force on the CNT flexural coordinate (Figure~\ref{fig1}) \cite{sazonova_tunable_2004, garcia-sanchez_mechanical_2007}. Writing the mechanical displacement operator as $\hat{x}=x_{\rm zpf}(a+a^\dagger)$, where \(x_{\rm zpf}\) is the zero-point displacement amplitude, the~laboratory-frame drive Hamiltonian may be written as~\cite{palyi_spinorbitinduced_2012, chang_quantum_2025}
\vspace{-3pt}
\begin{equation}
H_d^{\rm lab}(t)=-F_d(t)\hat{x}
=-F_0x_{\rm zpf}(a+a^\dagger)\cos(\omega_d t),
\label{eq:force_drive}
\end{equation}
where \(F_d(t)=F_0\cos(\omega_d t)\) is the externally applied drive force acting on the CNT flexural~coordinate.

Following earlier work on suspended CNT spin qubits, spin--motion coupling can arise when the nanotube displacement modulates the local magnetic field seen by the spin or alters a spin--orbit mediated quantization axis~\cite{palyi_spinorbitinduced_2012, wang_creating_2016}. In~the fullerene setting considered in Figure~\ref{fig1}, an~externally imposed magnetic field gradient is especially natural: if the fullerene spin is displaced together with the nanotube, then its Zeeman splitting becomes position dependent. Quantizing the displacement yields a spin--phonon interaction linear in the oscillator coordinate. Near~resonance and after a rotating-wave approximation, this reduces to a Jaynes--Cummings type coupling, as~we show in the next~section.

%%%%%%%%%%%%%%%%%%%%%%%%%%%%%%%%%%%%%%%%%%
\section{Hybrid Open-System Model}
\label{sec:model}

\subsection{System Hamiltonian}

The total Hamiltonian is written as
\vspace{-3pt}
\begin{equation}
H(t)=H_s+H_m+H_{\rm int}+H_d^{\rm lab}(t),
\end{equation}
with
\begin{equation}
H_s=\frac{\hbar\omega_s}{2}\sigma_z,
\qquad
H_m=\hbar\omega_m a^\dagger a,
\end{equation}
and with the effective spin--phonon interaction:
\vspace{-3pt}\begin{equation}
H_{\rm int}=\hbar g_1\left(a+a^\dagger\right)\sigma_x,
\label{eq:Hintfull}
\end{equation}
where $g_1$ is the displacement-mediated coupling strength. Here the Pauli operators are defined on the $\{|g\rangle,|e\rangle\}$ manifold as
$\sigma_z=|e\rangle\!\langle e|-|g\rangle\!\langle g|$ and
$\sigma_x=|g\rangle\!\langle e|+|e\rangle\!\langle g|$.

Using $\sigma_x=\sigma_++\sigma_-$, one may decompose $H_{\rm int}$ into resonant and counter-rotating contributions. Close to resonance, $\omega_s\approx\omega_m$, and~for $g_1\ll \omega_s,\omega_m$, the~rotating-wave approximation (RWA) neglects the rapidly oscillating counter-rotating terms $a\sigma_-$ and $a^\dagger\sigma_+$. Moving to a frame rotating at $\omega_m$ and applying RWA yields the Jaynes--Cummings Hamiltonian (see Appendix~\ref{appA})
\vspace{-3pt}\begin{equation}
H_{\rm JC}(t)=
\frac{\hbar\Delta}{2}\sigma_z
+\hbar g_1\left(a\sigma_+ + a^\dagger\sigma_-\right)
-\frac{F_0x_{\rm zpf}}{2}(a+a^\dagger).
\label{eq:HJC}
\end{equation}
where $\Delta=\omega_s-\omega_m$ is the detuning frequency, and~$\sigma_+=|e\rangle\!\langle g|$ , $\sigma_-=|g\rangle\!\langle e|$ are the spin ladder operators. Equation~\eqref{eq:HJC} describes coherent exchange between the fullerene qubit and the CNT phonon~mode.

In practice, the~coupling $g_1$ may contain contributions from intrinsic spin--orbit--deflection mechanisms and from externally generated magnetic gradients. A~useful estimate for the gradient-induced term is~\cite{palyi_spinorbitinduced_2012}
\vspace{-3pt}\begin{equation}
g_1 \sim \frac{g_e\mu_B}{\hbar}\left(\frac{\partial B}{\partial x}\right)x_{\rm zpf},
\label{eq:g1}
\end{equation}
where $g_e$ is the electronic $g$--factor, $\mu_B$ the Bohr magneton, $\partial B/\partial x$ the local gradient, and~$x_{\rm zpf}$ the zero-point fluctuation amplitude of the CNT flexural mode. This expression makes explicit the central design principle: increasing the magnetic gradient or softening the mechanical mode enhances the spin--vibrational~hybridization.

\subsection{Dissipative channels and Lindblad equation}

The hybrid system interacts with external environments through both spin and mechanical channels. We model the reduced density operator $\rho$ of the coupled spin--mechanical system with the Lindblad master equation~\cite{schlosshauer_decoherence_2007, breuer_theory_2009}
\vspace{-3pt}\begin{equation}
\dot\rho = -\frac{i}{\hbar}[H_{\rm JC}(t),\rho] + \mathcal{L}_m[\rho] + \mathcal{L}_s[\rho].
\label{eq:lme}
\end{equation}
For the mechanical mode we take the dissipator term~\cite{schlosshauer_decoherence_2007, breuer_theory_2009, palyi_spinorbitinduced_2012}.
\begin{align}
\mathcal{L}_m[\rho] 
&\;=\; \gamma_m(\bar n_m+1)\left(a\rho a^{\dagger} - \frac{1}{2}\{a^{\dagger}a,\rho\}\right) \nonumber \\
 &\;+\;
 \gamma_m\bar n_m\left(a^{\dagger}\rho a - \frac{1}{2}\{aa^{\dagger},\rho\}\right),
\label{eq:Lm}
\end{align}
where $\gamma_m$ is the mechanical damping rate and $\bar n_m = [\exp(\hbar\omega_m/k_BT_m)-1]^{-1}$ is the thermal phonon occupation of the mechanical~reservoir.

For the spin qubit we include relaxation, thermal excitation, and~pure dephasing, and~the dissipator is given by~\cite{schlosshauer_decoherence_2007, breuer_theory_2009, chang_quantum_2025}
\begin{align}
&\mathcal{L}_s[\rho] =
\Gamma_{\downarrow}\left(\sigma_-\rho\sigma_+  
- \frac{1}{2}\{\sigma_+\sigma_-,\rho\}\right) \, +\nonumber\\
&{}+
\Gamma_{\uparrow}\left(\sigma_+\rho\sigma_- 
- \frac{1}{2}\{\sigma_-\sigma_+,\rho\}\right)
+
\frac{\Gamma_\phi}{2}\left(\sigma_z\rho\sigma_z - \rho\right).
\label{eq:Ls}
\end{align}

The rates $\Gamma_{\downarrow}$ and $\Gamma_{\uparrow}$ denote spin relaxation and thermal excitation, respectively, and~are determined by the bath spectral density at the spin transition frequency $\omega_s$ \cite{chang_quantum_2025}. The~rate $\Gamma_\phi$ accounts for pure dephasing induced by, for~example, magnetic noise or slow charge fluctuations that shift the spin splitting. In~this effective description the spin rates are not derived from a single microscopic bath model. Rather, \(\Gamma_{\downarrow}\), \(\Gamma_{\uparrow}\), and~\(\Gamma_{\phi}\) are phenomenological rates that summarize spin relaxation, thermally activated excitation, and~pure dephasing due to several possible microscopic mechanisms, including phonon mediated relaxation in the CNT/fullerene environment, magnetic field noise from the device and micromagnet, charge-noise induced shifts of the local spin splitting, and~nuclear-spin or hyperfine-related spectral diffusion. They can be related to experimentally measured spin-relaxation and coherence times through $1/T_1=\Gamma_{\downarrow}+\Gamma_{\uparrow}\equiv\Gamma_1$, and~$1/T_2=(\Gamma_{\downarrow}+\Gamma_{\uparrow})/2+\Gamma_{\phi}\equiv \Gamma_2$ \cite{schlosshauer_decoherence_2007, breuer_theory_2009, chang_quantum_2025}. 

For thermal spin reservoirs, the~excitation and relaxation rates obey detailed balance, \(\Gamma_{\uparrow}/\Gamma_{\downarrow}=\exp[-\hbar\omega_s/(k_BT_s)]\), so that \(\Gamma_{\uparrow}\) is strongly suppressed at low temperature when \(\hbar\omega_s\gg k_BT_s\) \cite{schlosshauer_decoherence_2007, breuer_theory_2009}. Literature values for endohedral fullerene spins give coherence times in the range \(T_2\sim 10^{-5}\)--\(10^{-4}\,{\rm s}\), and~up to \(T_2\simeq 0.2\)--\(0.3\,{\rm ms}\) for \(N@C_{60}\) under favorable conditions~\cite{morton_relaxation_2006, morton_environmental_2007}. Encapsulation in a CNT device can also modify the spin-lattice relaxation {environment}%MDPI: References should be numbered in order of appearance. We detected "Ref 52" appears before "46--51", so we rearranged all the references to appear in numerical order, please check and confirm.
~\cite{harneit_fullerenes_2017, toth_enhanced_2008}. Therefore, in~a suspended CNT device with nearby gates and a micromagnet, additional magnetic and charge noise may increase the effective dephasing rate, and~the representative value \(\Gamma_2=\Gamma_{\phi}+(\Gamma_{\downarrow}+\Gamma_{\uparrow})/2\sim 10^5\,{\rm s}^{-1}\) used below should be viewed as a conservative device-level estimate rather than an intrinsic fullerene limit.
Equations~\eqref{eq:lme}--\eqref{eq:Ls} define the dynamical generator used throughout this work. They provide a minimal but flexible description of entropy generation in the driven hybrid~system.

\subsection{Reduced equations and weak-coupling structure}

The full density operator contains both spin and oscillator correlations. For~analytical insight, it is useful to identify two complementary limits. In~the weak-hybridization regime $g_1 \ll \gamma_m,\Gamma_2$, the~mechanical mode behaves as a damped driven oscillator~\cite{chang_quantum_2025} that perturbs the spin through a structured noise spectrum. In~the near-resonant strong cooperativity regime $g_1$ competes with the dissipative rates, so that coherent excitation exchange appreciably reshapes the entropy budget. The~Wigner-function formulation developed below is especially convenient in the first regime, but~it also remains useful in the second as a way to visualize the oscillator's contribution to the joint~state.

\section{Wigner-Function Description of the CNT Resonator}
\label{sec:wigner}

\subsection{Driven quantum Brownian motion and exact propagator}

In the absence of coupling to the spin, the~CNT flexural mode reduces to a damped driven quantum oscillator~\cite{chang_quantum_2025}. The~Wigner-function formalism provides a particularly useful phase-space representation for this mode because it encodes the full information contained in the oscillator density matrix while offering an intuitive picture of the dynamics in terms of quasi-probability flow in position--momentum space~\cite{weinbub_recent_2018, case_wigner_2008}. Unlike a classical probability distribution, the~Wigner function can take negative values, thereby encoding genuinely quantum features such as interference and nonclassicality, while still allowing damping, diffusion, and~coherent driving to be visualized in a form closely connected to classical trajectories~\cite{bertet_direct_2002, lougovski_fresnel_2003, karlovets_possibility_2017, chang_quantum_2025}. For~linear systems coupled to Gaussian environments, this formalism is especially powerful because the time evolution of the mechanical state can be expressed exactly through a Gaussian propagator, with~the deterministic drift of the phase-space center and the noise-induced broadening of the distribution appearing separately and transparently~\cite{PhysRevD.64.105020, qiu_quantum_2021}. This makes the Wigner representation particularly well suited for the present problem, where we wish to distinguish reversible transport from irreversible diffusion and damping, and~later relate these features directly to entropy flow and entropy~production.

In phase space, the~reduced mechanical state is represented by the Wigner function~\cite{case_wigner_2008}
\begin{equation}
W(q,p,t) = \frac{1}{2\pi\hbar}\int_{-\infty}^{\infty} d\xi\; e^{-ip\xi/\hbar}\left\langle q+\frac{\xi}{2}\right|\rho_m(t)\left|q-\frac{\xi}{2}\right\rangle,
\end{equation}
where $q$ and $p$ denote the canonical position and momentum variables of the CNT flexural mode in phase space, and~$\rho_m$ is the oscillator density matrix. For~linear damping and Gaussian reservoirs, the~Wigner function evolves under an exact Gaussian propagator of the form~\cite{PhysRevD.64.105020}
\begin{equation}
W(\mathbf{x},t) = \int d^2\mathbf{x}'\; K_W(\mathbf{x},t|\mathbf{x}',0)\,W(\mathbf{x}',0),
\label{eq:Wprop}
\end{equation}
with
 $\mathbf{x}=(q,p)^T$. The~propagator can be written schematically as~\cite{PhysRevD.64.105020}
\begin{align}
&K_W(\mathbf{x},t|\mathbf{x}',0)= \nonumber \\ 
&\frac{1}{2\pi\sqrt{\det\Sigma(t)}}
\exp\left[
-\frac{1}{2}
\left(\mathbf{x}-\bar{\mathbf{x}}(t;\mathbf{x}')\right)^T
\Sigma^{-1}(t)
\left(\mathbf{x}-\bar{\mathbf{x}}(t;\mathbf{x}')\right)
\right].
\label{eq:KWgaussian}
\end{align}
where $\bar{\mathbf{x}}(t;\mathbf{x}')$ is the classical damped driven trajectory launched from $\mathbf{x}'$ and $\Sigma(t)$ is the covariance matrix generated by environmental noise. This structure expresses the solution as a classical flow together with diffusive broadening~\cite{PhysRevD.64.105020, qiu_quantum_2021}.

For a harmonic oscillator of mass $m$ and frequency $\omega_m$ with a damping kernel reducible to a Markovian rate $\gamma_m$, the~phase-space center obeys
\begin{equation}
\dot q = \frac{p}{m}, \qquad \dot p = -m\omega_m^2 q - \gamma_m p + F_d(t),
\label{eq:classicaldrift}
\end{equation}
where $F_d(t)$ is the applied drive force (Equation~\eqref{eq:force_drive}). In~the Markovian limit of quantum Brownian motion, the~first moments obey damped-oscillator equations, while the second moments form a closed linear system that can be written in covariance-matrix form,
\vspace{-3pt}\begin{equation}
\dot\Sigma = A\Sigma + \Sigma A^T + D,
\label{eq:Sigmaeq}
\end{equation}
with drift matrix $A$ and diffusion matrix $D$ determined by the damping and bath temperature~\cite{breuer_theory_2009, PhysRevD.64.105020}.

This exact solution is especially useful because it cleanly separates three ingredients: \linebreak (i) reversible phase-space transport generated by the oscillator Hamiltonian and external drive; (ii) irreversible contraction associated with damping; (iii) noise-induced spreading governed by $D$. These ingredients map naturally onto the entropy balance discussed later in this paper. The~explicit expressions for the covariance matrix $\Sigma(t)$ and the mean phase-space trajectory $\bar{\mathbf{x}}(t)$, which determine the Gaussian propagator in \mbox{Equations~\eqref{eq:Wprop} and \eqref{eq:KWgaussian}}, are derived in Appendix~\ref{appB}.

\subsection{Embedding the propagator in the spin--phonon problem}

When the spin--phonon coupling is restored, the~oscillator no longer evolves autonomously. Nevertheless, the~exact oscillator propagator remains useful in two ways. First, in~perturbative treatments, the spin can enter as a source term or as a conditioned displacement, so that the oscillator Wigner function becomes a weighted superposition of Gaussian packets associated with spin-conditioned trajectories. Second, even in the fully coupled regime, the~reduced mechanical state
\vspace{-3pt}\begin{equation}
\rho_m(t) = \mathrm{Tr}_s\,\rho(t)
\end{equation}
can be monitored through its Wigner function to diagnose the onset of irreversibility: phase-space broadening, loss of negativity, and~displacement diffusion provide useful indicators of increasing decoherence and entropy~production.

Writing the joint density operator in the spin basis,
\begin{equation}
\rho(t) = \sum_{\alpha,\beta\in\{g,e\}} \rho_{\alpha\beta}^{(m)}(t) \otimes |\alpha\rangle\langle\beta|,
\end{equation}
one can define a matrix-valued Wigner function,
\begin{equation}
\mathbb{W}(q,p,t) =
\begin{pmatrix}
W_{gg}(q,p,t) & W_{ge}(q,p,t) \\
W_{eg}(q,p,t) & W_{ee}(q,p,t)
\end{pmatrix},
\end{equation}
whose diagonal components describe oscillator phase-space distributions conditioned on the spin state, and~whose off-diagonal components encode spin--mechanical coherence. The~Lindblad equation induces coupled partial differential equations for these entries. In~the weak-coupling limit, the~diagonal components can remain approximately Gaussian, while the off-diagonal terms typically decay on a scale set by $\Gamma_2$ together with coupling-induced and mechanical~dephasing. 

Experimentally, reconstruction of the matrix-valued Wigner function would require joint spin--mechanical tomography rather than tomography of the reduced mechanical state alone. The~diagonal components \(W_{gg}(q,p,t)\) and \(W_{ee}(q,p,t)\) could in principle be reconstructed by repeating the experiment many times, measuring the PEF spin state in the \(\{|g\rangle,|e\rangle\}\) basis, and~sorting the subsequent mechanical tomography data according to whether the spin was found in \(|g\rangle\) or \(|e\rangle\) \cite{chang_quantum_2025, bertet_direct_2002, lougovski_fresnel_2003, karlovets_possibility_2017, leibfried_experimental_1996}. 
The off-diagonal components \(W_{ge}(q,p,t)\) and \(W_{eg}(q,p,t)\), which encode spin--mechanical coherence, would require measurements in rotated spin bases. This could be implemented using Ramsey-type \(\pi/2\) pulses before spin readout, thereby mapping the real and imaginary parts of the spin coherence onto population differences in the \(\sigma_x\) and \(\sigma_y\) bases~\cite{chang_quantum_2025, leibfried_experimental_1996}. Such a protocol would therefore require independent spin initialization, coherent spin rotations, and~spin-selective readout, together with mechanical displacement and parity measurement. In~practice, the~spin measurement would introduce backaction by projecting the joint spin--mechanical state onto a conditional mechanical state. Thus, the reconstructed Wigner matrix should be understood as an ensemble tomographic object obtained from repeated preparations, not from a single nondestructive measurement~\cite{chang_quantum_2025, case_wigner_2008, bertet_direct_2002, lougovski_fresnel_2003, karlovets_possibility_2017, leibfried_experimental_1996}. The~entropy production analysis in the present work refers to the premeasurement open system dynamics. If~the measurement process itself is included, the~spin-readout backaction and information gain would constitute an additional measurement-induced entropy and information flow channel, which could be incorporated using a quantum-trajectory or generalized measurement extension of the Lindblad description~\cite{breuer_theory_2009}.

\subsection{Mechanical entropy in phase space}

The von Neumann entropy of the reduced mechanical state, $S_m = -\mathrm{Tr}(\rho_m\ln\rho_m)$, generally cannot be written exactly as a simple functional of the Wigner function because $W(q,p)$ need not be positive. However, for~a single-mode Gaussian state, the entropy is completely determined by the symplectic eigenvalue of the covariance matrix $\Sigma$ \cite{breuer_theory_2009}. Defining
\begin{equation}
\nu = \frac{1}{\hbar}\sqrt{\det(2\Sigma)},
\end{equation}
with $\nu \ge 1$, one may identify an effective thermal occupation number through $\nu = 2\bar n + 1$, or~equivalently $\bar n = (\nu - 1)/2$. Because~any single-mode Gaussian state is unitarily equivalent to a thermal state with this occupation number, its entropy is
\begin{align}
& S_m = -\mathrm{Tr}(\rho_m\ln\rho_m)   \nonumber\\
& =\left(\frac{\nu+1}{2}\right)\ln\left(\frac{\nu+1}{2}\right) - \left(\frac{\nu-1}{2}\right)\ln\left(\frac{\nu-1}{2}\right).
\label{eq:gaussentropy}
\end{align}
Equation~\eqref{eq:gaussentropy} makes the relation between the covariance matrix and the mechanical entropy explicit: for a single-mode Gaussian state, $S_m$ is a monotonic function of the symplectic eigenvalue $\nu$, and~hence of $\det\Sigma$. Accordingly, whenever diffusion and dissipation increase $\det\Sigma$, the~mechanical entropy also~increases.

\section{Entropy Balance, Entropy Flux, and Entropy Production}
\label{sec:entropy}

\subsection{Von Neumann entropy balance for Lindblad dynamics}
\label{sec:entropy_balance}

The total nonequilibrium entropy of the coupled spin--mechanical system is the von Neumann entropy
\begin{equation}
S(t)=-\mathrm{Tr}[\rho(t)\ln\rho(t)].
\end{equation}
The reduced density operator evolves according to the Lindblad master equation
\begin{equation}
\dot\rho(t)=\mathcal{L}\rho(t)=-\frac{i}{\hbar}[H(t),\rho(t)]+\sum_j \left(L_j\rho L_j^{\dagger}-\frac{1}{2}\{L_j^{\dagger}L_j,\rho\}\right)
\end{equation}
where $\mathcal{L}$ is the full dynamical generator and the $L_j$ are the jump operators \cite{breuer_theory_2009}.
Differentiating $S(t)$ with respect to time gives
\begin{equation}
\frac{dS}{dt}
=-\mathrm{Tr}[\dot\rho\ln\rho]
-\mathrm{Tr}\!\left[\rho\,\frac{d}{dt}(\ln\rho)\right].
\end{equation}
Since $\mathrm{Tr}\rho=1$ at all times, the second term vanishes, so that
\begin{equation}
\frac{dS}{dt}=-\mathrm{Tr}[(\mathcal{L}\rho)\ln\rho].
\label{eq:dSdt_basic}
\end{equation}
The Hamiltonian part of $\mathcal{L}$ does not contribute to $dS/dt$, because the trace of the commutator term vanishes by cyclicity of the trace and the fact that $\rho$ commutes with $\ln\rho$. Thus the entropy change is governed entirely by the dissipative part of the evolution.

To identify the irreversible contribution, we introduce a stationary reference state $\rho_{\rm ss}$ satisfying
\begin{equation}
\mathcal{L}\rho_{\rm ss}=0
\end{equation}
and define the quantum \textit{relative} entropy
\begin{equation}
S(\rho(t)\|\rho_{\rm ss})
=\mathrm{Tr}\!\left[\rho(t)\bigl(\ln\rho(t)-\ln\rho_{\rm ss}\bigr)\right].
\end{equation}
Differentiating with respect to time yields
\begin{equation}
\frac{d}{dt}S(\rho\|\rho_{\rm ss})
=\mathrm{Tr}\!\left[(\mathcal{L}\rho)\bigl(\ln\rho-\ln\rho_{\rm ss}\bigr)\right],
\end{equation}
which motivates the definition of the entropy-production rate
\begin{equation}
\Pi(t)\equiv
-\frac{d}{dt}S(\rho(t)\|\rho_{\rm ss})
=
-\mathrm{Tr}\!\left[(\mathcal{L}\rho(t))
\bigl(\ln\rho(t)-\ln\rho_{\rm ss}\bigr)\right].
\label{eq:Pi}
\end{equation}
Under the standard assumptions of complete positivity and stationarity, Spohn's inequality ensures that \cite{breuer_theory_2009}
\begin{equation}
\Pi(t)\ge 0.
\end{equation}

Thus $\Pi(t)$ measures the irreversible entropy generated by the open-system dynamics. Expanding Eq.~\eqref{eq:Pi} and using Eq.~\eqref{eq:dSdt_basic}, one obtains
\begin{equation}
\Pi(t)=\frac{dS}{dt}
+\mathrm{Tr}[(\mathcal{L}\rho)\ln\rho_{\rm ss}].
\end{equation}
This suggests defining the entropy-flux rate as
\begin{equation}
\Phi(t)\equiv
-\mathrm{Tr}\!\left[(\mathcal{L}\rho(t))\ln\rho_{\rm ss}\right],
\label{eq:Phi}
\end{equation}
so that the entropy balance takes the form
\begin{equation}
\frac{dS}{dt}=\Pi(t)-\Phi(t).
\label{eq:balance}
\end{equation}
For thermal reservoirs, $\Phi$ reduces to the usual heat flux form when $\rho_{\rm ss}$ is Gibbsian. In the driven CNT--PEF system, however, coherent forcing and multiple baths generally imply that the relevant stationary state is not a simple global Gibbs state, making Eq.~\eqref{eq:Phi} the more general definition. In a nonequilibrium steady state, $dS/dt=0$, and Eq.~\eqref{eq:balance} reduces to
\begin{equation}
\Pi_{\rm ss}=\Phi_{\rm ss}.
\end{equation}

\subsection{Mode-resolved decomposition}

Because the dissipator is the sum of spin and mechanical contributions, the~entropy production may be decomposed as~\cite{breuer_theory_2009}
\vspace{-3pt}\begin{equation}
\Pi=\Pi_m+\Pi_s+\Pi_{\rm corr},
\end{equation}
where $\Pi_m$ and $\Pi_s$ are associated with $\mathcal{L}_m$ and $\mathcal{L}_s$, respectively, and~$\Pi_{\rm corr}$ accounts for the fact that the logarithm of the full state does not generally decompose additively once spin–mechanical correlations develop. In~the weak-correlation regime one may~approximate
\vspace{-3pt}\begin{equation}
\rho\simeq \rho_s\otimes\rho_m,
\end{equation}
which yields
\begin{equation}
\Pi\simeq \Pi_s^{\rm red}+\Pi_m^{\rm red}.
\end{equation}
This decomposition is useful for identifying whether irreversibility is dominated by vibrational diffusion or by spin relaxation and dephasing~\cite{Landi:2020bsq}.

For a thermal mechanical bath with reference state
\begin{equation}
\rho_m^{\rm th}
=
\frac{e^{-\beta_m\hbar\omega_m a^\dagger a}}
{\mathrm{Tr}[e^{-\beta_m\hbar\omega_m a^\dagger a}]},
\end{equation}
one finds
\begin{equation}
\Phi_m
=
-\mathrm{Tr}[\mathcal{L}_m\rho\ln\rho_m^{\rm th}]
\simeq
\beta_m J_m,
\label{eq:fux}
\end{equation}
where $J_m$ denotes the rate of energy transfer from the mechanical mode to its thermal reservoir in the weak-correlation (factorized) limit. An~analogous expression applies to the spin bath when
$\Gamma_\uparrow/{\Gamma_\downarrow}=e^{-\beta_s\hbar\omega_s}$ \cite{Landi:2020bsq, breuer_theory_2009}.

\subsection{Entropy production in the Wigner representation}

For the oscillator sector, the~Wigner formulation permits a semiclassical entropy analysis when the state remains Gaussian or nearly Gaussian. In~the linear Markovian regime, the~phase space dynamics can be written in continuity-equation form~\cite{breuer_theory_2009},
\vspace{-3pt}\begin{equation}
\partial_t W = -\nabla\cdot(\mathbf{J}_{\rm rev} + \mathbf{J}_{\rm irr}),
\end{equation}
where $\mathbf{J}_{\rm rev}$ is the reversible current generated by Hamiltonian flow and $\mathbf{J}_{\rm irr}$ is the irreversible current arising from damping and diffusion. For~Gaussian states, whose Wigner functions are positive, one typically introduces the Shannon-like phase space entropy
\begin{equation}
S_W = -\int dq\,dp\; W(q,p,t)\ln W(q,p,t),
\end{equation}
which provides a useful coarse-grained measure of the oscillator's entropy. Although~$S_W$ is not, in~general, identical to the von Neumann entropy, for~Gaussian states it is closely related to it through the covariance matrix and therefore serves as an intuitive diagnostic of diffusion-driven irreversibility~\cite{Landi:2020bsq, schlosshauer_decoherence_2007, breuer_theory_2009}.

In this representation, the~irreversible phase-space current can be used to express entropy production. In~particular, for~linear Fokker--Planck dynamics, one obtains a quadratic form involving the drift and diffusion matrices, showing that entropy is generated whenever damping-induced contraction and thermal broadening fail to balance at the instantaneous state~\cite{schlosshauer_decoherence_2007, PhysRevD.64.105020}. The~exact propagator of Section~\ref{sec:wigner} therefore provides not only the state itself but also the ingredients needed to characterize the oscillator's entropy~budget.

\begin{figure*}[t!]
    \centering
    \includegraphics[width=0.95\linewidth]{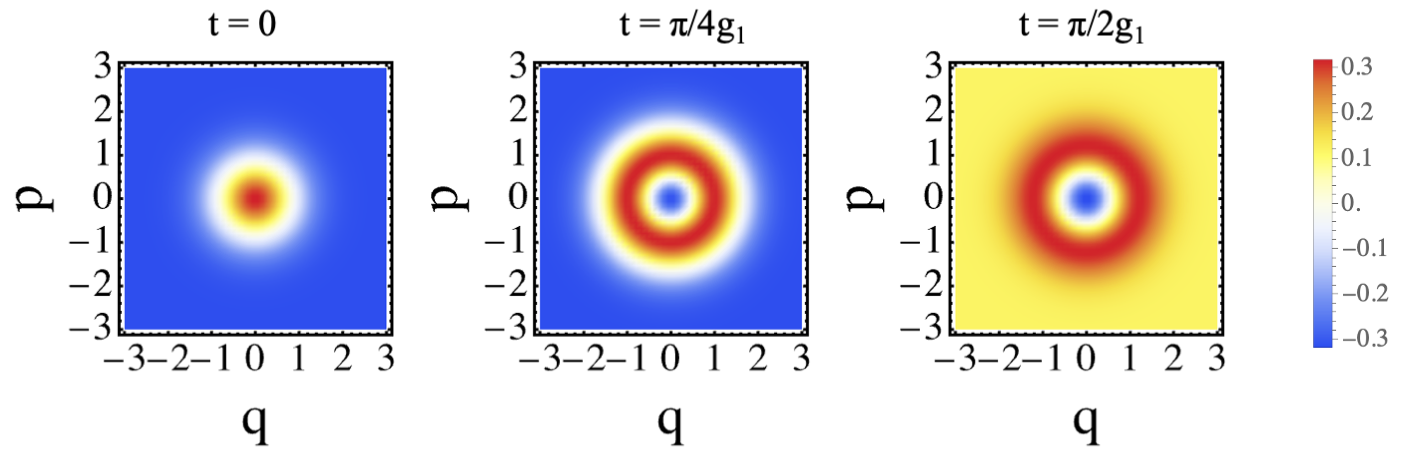}
    \caption{Reduced mechanical Wigner function $W_m(q,p,t)$ in the resonant weak-drive regime, evaluated from Eq.~\eqref{eq:Wweakdrive} at three representative times. At~$t=0$, the~oscillator is in the vacuum state and the Wigner function is purely Gaussian. At~intermediate times $t=\pi/(4g_1)$, the~phase-space distribution is a mixture of the vacuum and one-phonon contributions. At~$t=\pi/(2g_1)$, the~oscillator is in the one-phonon state, and~the Wigner function displays the characteristic nonclassical structure associated with a single vibrational excitation. This evolution provides a direct visualization of the coherent transfer of a single excitation from the spin sector to the CNT mode. The~color bar indicates the magnitude of $W_m(q,p,t)$.  \label{fig2}}
\end{figure*}  

\section{Representative Regimes and Entropy-Production Signatures}
\label{sec:results}

\subsection{Resonant weak-drive regime}

Consider first a near-resonant regime $\omega_s\approx\omega_m$ with weak coherent drive and low bath temperature, such that $\bar n_m\ll 1$ and $\Gamma_{\uparrow}\ll\Gamma_{\downarrow}$. If~the system is initialized in $|e,0\rangle$, coherent spin--phonon exchange produces Rabi-like oscillations between $|e,0\rangle$ and $|g,1\rangle$ \cite{chang_quantum_2025}. Here $\lvert 0\rangle$ and $\lvert 1\rangle$ denote the zero- and one-phonon Fock states of the CNT flexural mode, respectively, so that $\lvert e,0\rangle$ and $\lvert g,1\rangle$ represent product states of the spin and mechanical sectors within the single-excitation manifold. In~the absence of dissipation this exchange is reversible, and the entropy remains low if the state is pure. Once mechanical damping and spin dephasing are included, these oscillations are damped and entropy production becomes~positive.

A simple concrete example is obtained by neglecting the weak drive to leading order over a single exchange cycle and truncating the Hilbert space to the single-excitation manifold $\{|e,0\rangle,|g,1\rangle\}$. Under~the resonant Jaynes--Cummings Hamiltonian, the~pure state evolves as~\cite{schlosshauer_decoherence_2007}
\begin{equation}
|\psi(t)\rangle = \cos(g_1t)\,|e,0\rangle - i\sin(g_1t)\,|g,1\rangle.
\label{eq:psiweakdrive}
\end{equation}
Tracing over the spin gives the reduced mechanical state~\cite{schlosshauer_decoherence_2007, breuer_theory_2009}
\begin{equation}
\rho_m^{(0)}(t)=\cos^2(g_1t)\,|0\rangle\langle 0|+\sin^2(g_1t)\,|1\rangle\langle 1|,
\label{eq:rhomweakdrive}
\end{equation}
so that the mechanical entropy is (Eqn. 22):
\begin{equation}
S_m^{(0)}(t)
=-\cos^2(g_1t)\ln[\cos^2(g_1t)]-\sin^2(g_1t)\ln[\sin^2(g_1t)].
\label{eq:Smweakdrive}
\end{equation}

Equation~\eqref{eq:Smweakdrive} shows explicitly that the reduced mechanical entropy 
oscillates between $0$ and $\ln 2$, vanishing at $t=n\pi/2g_1$ and reaching its maximum at $g_1t=\pi/4+\pi n/2$. This behavior reflects reversible spin--phonon entanglement generated by coherent Jaynes--Cummings dynamics, rather than irreversible entropy~production. 

The corresponding reduced mechanical Wigner function also follows directly from Equation~\eqref{eq:rhomweakdrive}. Since $\rho_m^{(0)}(t)$ is a mixture of the vacuum and one-phonon states (see \linebreak Appendix~\ref{appC}):
\begin{equation}
W_m(q,p,t)=\cos^2(g_1t)\,W_0(q,p)+\sin^2(g_1t)\,W_1(q,p),
\label{eq:Wweakdrive}
\end{equation}
where $W_0$ and $W_1$ are the Wigner functions of the harmonic-oscillator ground state and first excited state, respectively. Thus, the phase-space distribution oscillates between a purely Gaussian profile and a nonclassical one-phonon profile, providing a direct visualization of the coherent transfer of a single excitation from the spin sector to the CNT mode Figure~\ref{fig2}.

To connect this coherent dynamics to entropy production, we now include weak dissipation perturbatively. In~the low-temperature limit, the~leading mechanical heat current into the bath is proportional to the instantaneous phonon occupation~\cite{breuer_theory_2009}
\begin{equation}
J_m(t)\simeq \hbar\omega_m\gamma_m\,\langle a^\dagger a\rangle
=\hbar\omega_m\gamma_m\sin^2(g_1t),
\label{eq:Jmweakdrive}
\end{equation}
while the corresponding entropy flux is (Equation~\eqref{eq:fux}):
\begin{equation}
\Phi_m(t)\simeq \beta_m J_m(t)
=\beta_m\hbar\omega_m\gamma_m\sin^2(g_1t).
\label{eq:Phimweakdrive}
\end{equation}
Likewise, the~spin-bath channel contributes
\begin{equation}
J_s(t)\simeq \hbar\omega_s\Gamma_{\downarrow}\,\langle \sigma_+\sigma_-\rangle
=\hbar\omega_s\Gamma_{\downarrow}\cos^2(g_1t),
\label{eq:Jsweakdrive}
\end{equation}
and therefore
\begin{equation}
\Phi_s(t)\simeq \beta_s J_s(t)
=\beta_s\hbar\omega_s\Gamma_{\downarrow}\cos^2(g_1t),
\label{eq:Phisweakdrive}
\end{equation}
where we have neglected the thermally activated spin-excitation channel $\Gamma_\uparrow$.

\begin{figure*}
    \centering
    \includegraphics[width=0.8\linewidth]{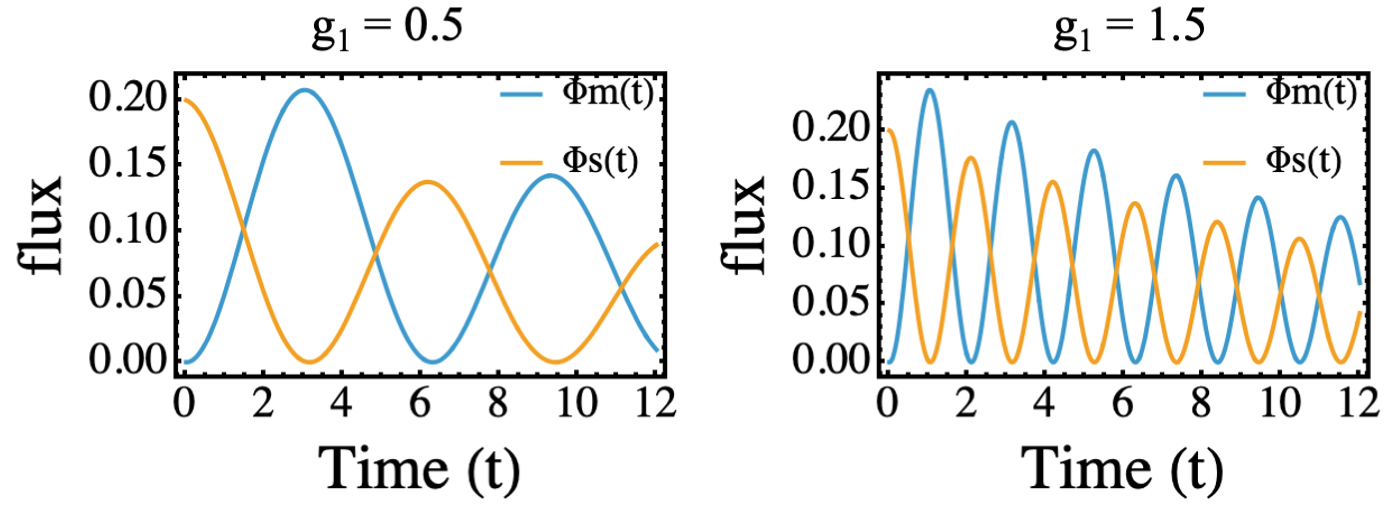}
    \caption{Entropy-flux dynamics in the resonant weak-drive regime for two representative values of the spin--phonon coupling $g_1$. The~mechanical entropy flux $\Phi_m(t)$ and spin entropy flux $\Phi_s(t)$ are plotted using the weak-excitation expressions derived in the text. As~$g_1$ increases, coherent spin--phonon exchange occurs on a shorter timescale, leading to a faster buildup of entropy flow through the mechanical channel and a corresponding redistribution of dissipation away from direct spin relaxation and toward the vibrational~mode. \label{fig3}}
\end{figure*}

\begin{figure*}[t!]
    \centering
    \includegraphics[width=0.9\linewidth]{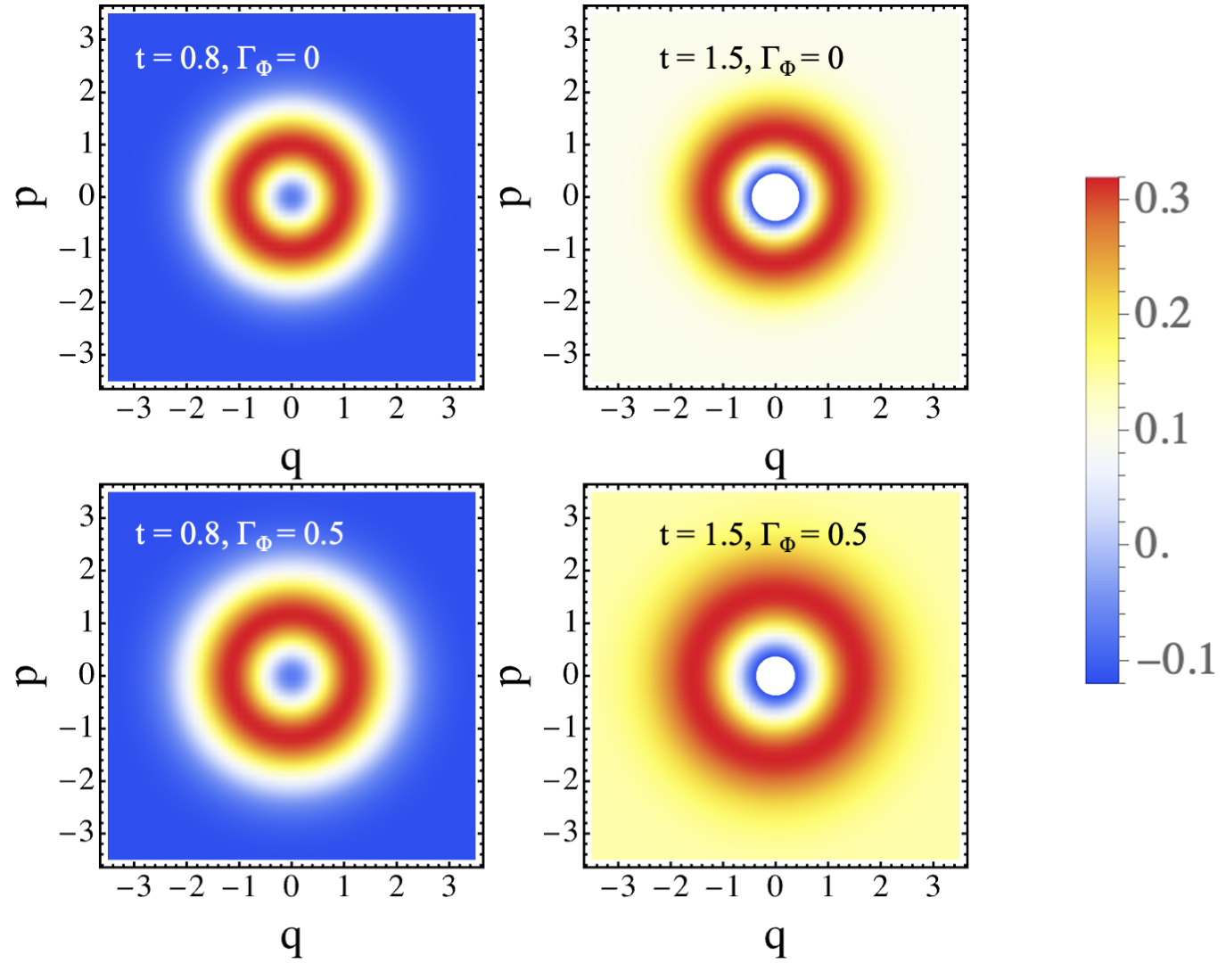}
    \caption{Diffusion-broadened reduced mechanical Wigner function $W_m(q,p,t)$ at representative times and dephasing rates in the resonant weak-drive regime. The~four panels show the phase-space distribution for two times, $t=\pi/(4g_1)$ and $t=\pi/(2g_1)$, and~for two values of the dephasing rate, $\Gamma_\phi=0$ and $\Gamma_\phi=0.5$. As~dephasing increases, the~nonclassical one-phonon structure is progressively smoothed and the distribution broadens in phase space, illustrating the combined effects of mechanical diffusion and spin decoherence. The~color bar indicates the magnitude of $W_m(q,p,t)$.   \label{fig4}}
\end{figure*}

\begin{figure*}
    \centering
    \includegraphics[width=0.8\linewidth]{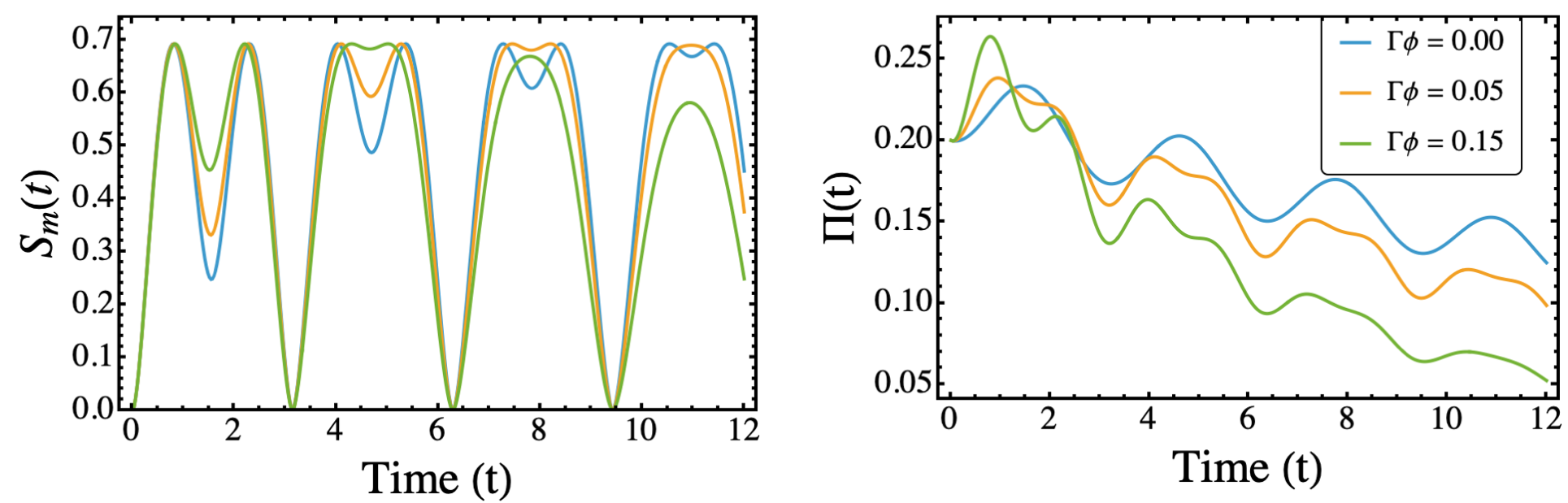}
    \caption{Left panel: reduced mechanical entropy $S_m(t)$ for three representative dephasing rates $\Gamma_\phi$ in the resonant weak-drive regime. Increasing dephasing progressively damps the oscillatory behavior associated with coherent spin--phonon exchange. Right panel: corresponding irreversible contribution $\Pi(t)$ obtained from the phenomenological weak-excitation model. As~$\Gamma_\phi$ increases, the~entropy balance acquires a more pronounced strictly positive irreversible component arising from the combined effects of mechanical dissipation and spin decoherence. The~legend shown in the inset of the right panel applies to both~figures. \label{fig5}}
\end{figure*}

These expressions make the redistribution of irreversibility transparent. At~very short~times,
\begin{equation}
\Phi_m(t)\approx \beta_m\hbar\omega_m\gamma_m g_1^2 t^2,
\qquad
\Phi_s(t)\approx \beta_s\hbar\omega_s\Gamma_{\downarrow}(1-g_1^2 t^2),
\label{eq:shorttimefluxes}
\end{equation}
showing that the mechanical entropy flux turns on quadratically as the coherent exchange transfers excitation from the spin to the CNT mode. Averaging over one Rabi period $T_R=\pi/g_1$ gives
\begin{align}
&\overline{\Phi}_m
=\frac{1}{T_R}\int_0^{T_R}dt\,\Phi_m(t)
=\frac{1}{2}\beta_m\hbar\omega_m\gamma_m,  \nonumber \\
&\overline{\Phi}_s
=\frac{1}{2}\beta_s\hbar\omega_s\Gamma_{\downarrow}.
\label{eq:avgfluxes}
\end{align}
Thus, even in this simplest regime, the~formalism (Figure~\ref{fig3}) shows explicitly how increasing $g_1$ accelerates the transfer of excitation into the mechanical channel and thereby shifts part of the entropy flow from direct spin relaxation to dissipation through the vibrational mode. In~the presence of weak dephasing, the~oscillatory behavior in Equations~\eqref{eq:Smweakdrive}--\eqref{eq:Phisweakdrive} is progressively damped, and~the Wigner function broadens (Figure~\ref{fig4}). In~addition the entropy balance acquires a strictly positive irreversible component $\Pi(t)$ associated with both mechanical diffusion and spin decoherence (Figure~\ref{fig5}).

\subsection{Strong driving and phase-space distortion}

Under resonant or near-resonant coherent drive, the~oscillator acquires a large displacement, and the mechanical Wigner function is dragged along a damped orbit in phase space. In~the uncoupled case the state remains Gaussian, and the entropy increase is governed primarily by the covariance matrix $\Sigma(t)$. With~spin--phonon coupling, the~oscillator trajectory becomes spin conditioned, producing partial splitting or distortion of the Wigner distribution. These distortions reflect entangling dynamics and are accompanied by mutual information buildup between spin and~oscillator.

A simple concrete example can be obtained in the rotating frame of a resonant mechanical drive, $\omega_d=\omega_m$, in~the regime $\varepsilon\equiv F_0x_{\mathrm{zpf}}/\hbar\gg g_1$, where $\varepsilon$ sets the strength of the external coherent drive acting on the oscillator (Equation~\eqref{eq:force_drive}), and~for times short compared with the spin-relaxation time, so that the spin may be treated as approximately frozen in the $\sigma_x$ basis. In~that basis, the~interaction $\hbar g_1(a+a^\dagger)\sigma_x$ shifts the effective drive seen by the oscillator. If~the spin is initially prepared in the state $|e\rangle$, which can be written in the $\sigma_x$ basis as
\begin{equation}
|e\rangle=\frac{1}{\sqrt{2}}\left(|+\rangle_x+|-\rangle_x\right),
\end{equation}
and the mechanical mode starts in its ground state, then to leading order the joint state evolves into a spin-conditioned superposition of displaced oscillator states,
\begin{equation}
|\Psi(t)\rangle = \frac{1}{\sqrt{2}}\left(|+\rangle_x\,|\alpha_+(t)\rangle + |-\rangle_x\,|\alpha_-(t)\rangle\right),
\label{eq:strongdrivePsi}
\end{equation}
In the above equations $|+\rangle_x$ and $|-\rangle_x$ are the eigenstates of $\sigma_x$, related to the $\{|e\rangle,|g\rangle\}$ basis~by
\begin{equation}
|+\rangle_x=\frac{1}{\sqrt{2}}\left(|e\rangle+|g\rangle\right),
\qquad
|-\rangle_x=\frac{1}{\sqrt{2}}\left(|e\rangle-|g\rangle\right)
\end{equation}
In this basis the interaction is diagonal, so the strong-drive dynamics are naturally expressed in terms of spin-conditioned oscillator displacements. Here $|\alpha_+(t)\rangle$ and $|\alpha_-(t)\rangle$ are coherent states of the CNT flexural mode with spin-conditioned complex amplitudes $\alpha_\pm(t)$.

The coherent amplitudes obey
\begin{equation}
\dot\alpha_\pm(t)=-\frac{\gamma_m}{2}\alpha_\pm(t)-i\bigl(\varepsilon\pm g_1\bigr),
\qquad
\alpha_\pm(0)=0.
\label{eq:alphapmEq}
\end{equation}

For a constant real resonant drive amplitude $\varepsilon$, this gives
\begin{equation}
\alpha_\pm(t)=-\frac{2i}{\gamma_m}\bigl(\varepsilon\pm g_1\bigr)\left(1-e^{-\gamma_m t/2}\right).
\label{eq:alphapmSol}
\end{equation}
The mean oscillator displacement is therefore large when $|\varepsilon|/\gamma_m\gg 1$, while the spin-dependent splitting is controlled by
\begin{equation}
\delta\alpha(t)\equiv \alpha_+(t)-\alpha_-(t)
=-\frac{4ig_1}{\gamma_m}\left(1-e^{-\gamma_m t/2}\right).
\label{eq:deltaalpha}
\end{equation}

\begin{figure*}[t!]
    \centering
    \includegraphics[width=0.8\linewidth]{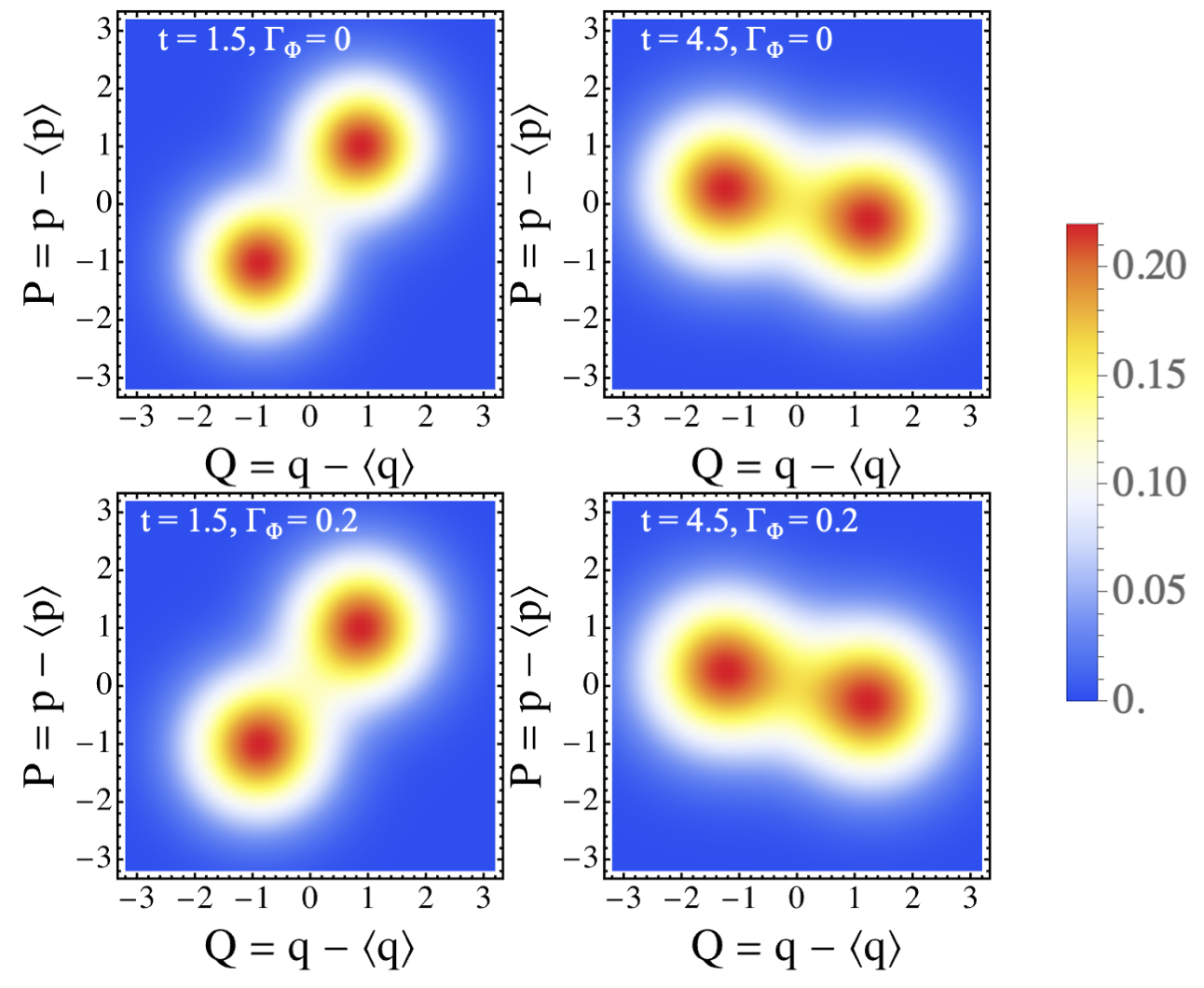}
    \caption{Diffusion-broadened reduced mechanical Wigner function in the strong-drive regime, shown in phase-space coordinates centered on the mean driven displacement, $Q=q-\langle q\rangle$ and $P=p-\langle p\rangle$. The~four panels display the phase-space distribution at two representative times and for two values of the dephasing rate $\Gamma_\phi$. As~time increases and dephasing is turned on, the~two spin-conditioned Gaussian packets broaden through the covariance matrix $\Sigma(t)$ and become increasingly distorted, while their overlap is reduced by decoherence. This evolution illustrates the crossover from reversible spin--oscillator entanglement dynamics to genuine irreversibility, with~a growing positive entropy-production contribution associated with both finite temperature diffusion and the generation and decay of spin--oscillator~correlations. \label{fig6}}
\end{figure*}

Tracing over the spin gives the reduced mechanical state
\begin{equation}
\rho_m(t)=\frac{1}{2}\Bigl(|\alpha_+(t)\rangle\langle\alpha_+(t)|+|\alpha_-(t)\rangle\langle\alpha_-(t)|\Bigr),
\label{eq:rhomStrongDrive}
\end{equation}
so that the mechanical Wigner function becomes a sum of two Gaussian packets with common covariance matrix $\Sigma(t)$ but different centers,
\begin{equation}
W_m(\mathbf{x},t)=\frac{1}{2}\,G\!\left(\mathbf{x}-\bar{\mathbf{x}}_+(t),\Sigma(t)\right)
+\frac{1}{2}\,G\!\left(\mathbf{x}-\bar{\mathbf{x}}_-(t),\Sigma(t)\right),
\label{eq:Wstrongdrive}
\end{equation}
where
\begin{equation}
G(\mathbf{x},\Sigma)=\frac{1}{2\pi\sqrt{\det\Sigma}}
\exp\!\left(-\frac{1}{2}\mathbf{x}^T\Sigma^{-1}\mathbf{x}\right)
\end{equation}
and $\bar{\mathbf{x}}_\pm(t)$ are the phase-space centers associated with $\alpha_\pm(t)$. Equation~\eqref{eq:Wstrongdrive} makes explicit how strong driving produces a large overall displacement, while the spin--phonon coupling generates a conditional splitting of the phase-space distribution. When the separation $|\delta\alpha(t)|$ becomes comparable to or larger than the Gaussian width set by $\Sigma(t)$, the~Wigner function develops a resolved double-peak structure. For~smaller separations, it appears as a broadened, distorted single packet (Figure~\ref{fig6}).

In the absence of diffusion, the~entangling part of the dynamics can be quantified analytically. The~overlap of the two coherent states is
\begin{equation}
\langle \alpha_-(t)|\alpha_+(t)\rangle
=
\exp\!\left[-\frac{|\delta\alpha(t)|^2}{2}+i\,\mathrm{Im}\!\bigl(\alpha_+ \alpha_-^*\bigr)\right],
\label{eq:cohoverlap}
\end{equation}
so the nonzero eigenvalues of $\rho_m(t)$ are
\begin{equation}
\lambda_\pm(t)=\frac{1}{2}\left(1\pm e^{-|\delta\alpha(t)|^2/2}\right).
\label{eq:lambdapmStrong}
\end{equation}
The reduced mechanical entropy is therefore
\begin{equation}
S_m(t)=-\lambda_+(t)\ln\lambda_+(t)-\lambda_-(t)\ln\lambda_-(t).
\label{eq:SmStrongDrive}
\end{equation}
This entropy grows from zero as the two spin-conditioned trajectories separate in phase space. In~the purely coherent limit, Equation~\eqref{eq:SmStrongDrive} measures reversible entanglement between the spin and the oscillator. For~the pure bipartite state in Equation~\eqref{eq:strongdrivePsi}, the~mutual information is
\begin{equation}
I_{s:m}(t)=2S_m(t),
\label{eq:MIstrong}
\end{equation}
showing explicitly that phase-space splitting is accompanied by the buildup of spin--oscillator~correlations.

The entropy flux to the mechanical bath can also be estimated directly. In~the factorized weak-diffusion limit, the~mechanical heat current is approximately
\begin{equation}
J_m(t)\simeq \hbar\omega_m\gamma_m\,\bar n_{\rm osc}(t),
\label{eq:JmStrong}
\end{equation}
where the mean oscillator occupation is
\begin{align}
&\bar n_{\rm osc}(t)
=
\frac{1}{2}\Bigl(|\alpha_+(t)|^2+|\alpha_-(t)|^2\Bigr) \nonumber \\
&=
\frac{4(\varepsilon^2+g_1^2)}{\gamma_m^2}\left(1-e^{-\gamma_m t/2}\right)^2.
\label{eq:noscStrong}
\end{align}
The corresponding entropy flux is
\begin{equation}
\Phi_m(t)\simeq \beta_m J_m(t)
=
\beta_m\hbar\omega_m\gamma_m\,\bar n_{\rm osc}(t).
\label{eq:PhimStrong}
\end{equation}
Equations ~\eqref{eq:noscStrong} and ~\eqref{eq:PhimStrong} show explicitly that in the strong-drive regime the entropy flux into the mechanical bath scales as $\varepsilon^2/\gamma_m$ and can become large even when the spin subsystem remains close to its local stationary state. This is the oscillator-dominated entropy-production~regime.

Once diffusion and spin dephasing are restored, the~two Gaussian packets in Equation~\eqref{eq:Wstrongdrive} broaden through the covariance matrix $\Sigma(t)$, while the overlap in Equation~\eqref{eq:cohoverlap} is further suppressed by decoherence (Figure~\ref{fig6}). As~a result, the~oscillatory entanglement dynamics cross over into genuine irreversibility: the Wigner function becomes increasingly broadened and distorted, the~mutual information ceases to be purely reversible, and~the entropy balance acquires a strictly positive production term $\Pi(t)$ associated with both finite-temperature diffusion and the continuous generation and decay of spin--oscillator~correlations.

\subsection{Thermal crossover and scaling estimates}   
At higher temperatures, thermal excitation of the mechanical mode increases $\bar n_m$ and broadens the phase space distribution, so that the fullerene qubit interacts not with a nearly pure bosonic mode but with a thermally occupied one. Two effects then follow. First, the~coherent signature of spin--vibrational hybridization is progressively washed out as thermal fluctuations randomize the spin dependence. Second, the~total entropy production can initially increase with temperature since a larger set of dissipative transitions becomes thermally accessible and the irreversible phase space current correspondingly~grows. 

When the drive injects work continuously into the system, the~resulting nonequilibrium steady state typically violates global detailed balance even if each bath separately is thermal~\cite{Landi:2020bsq, breuer_theory_2009}. In~that case the stationary state supports a strictly positive entropy-production rate,
\begin{equation} 
\Pi_{\rm ss}=\Phi_{\rm ss}>0, 
\end{equation} 
because $dS/dt=0$ in the steady state. This quantity provides an experimentally relevant summary of irreversibility. In~the present platform, $\Pi_{\rm ss}$ can be tuned by the magnetic gradient (which changes $g_1$, see Equation~\eqref{eq:g1}), the~drive amplitude, the~bath temperature, and~the quality factor of the suspended~CNT. 

\begin{figure}
    \includegraphics[width=10 cm]{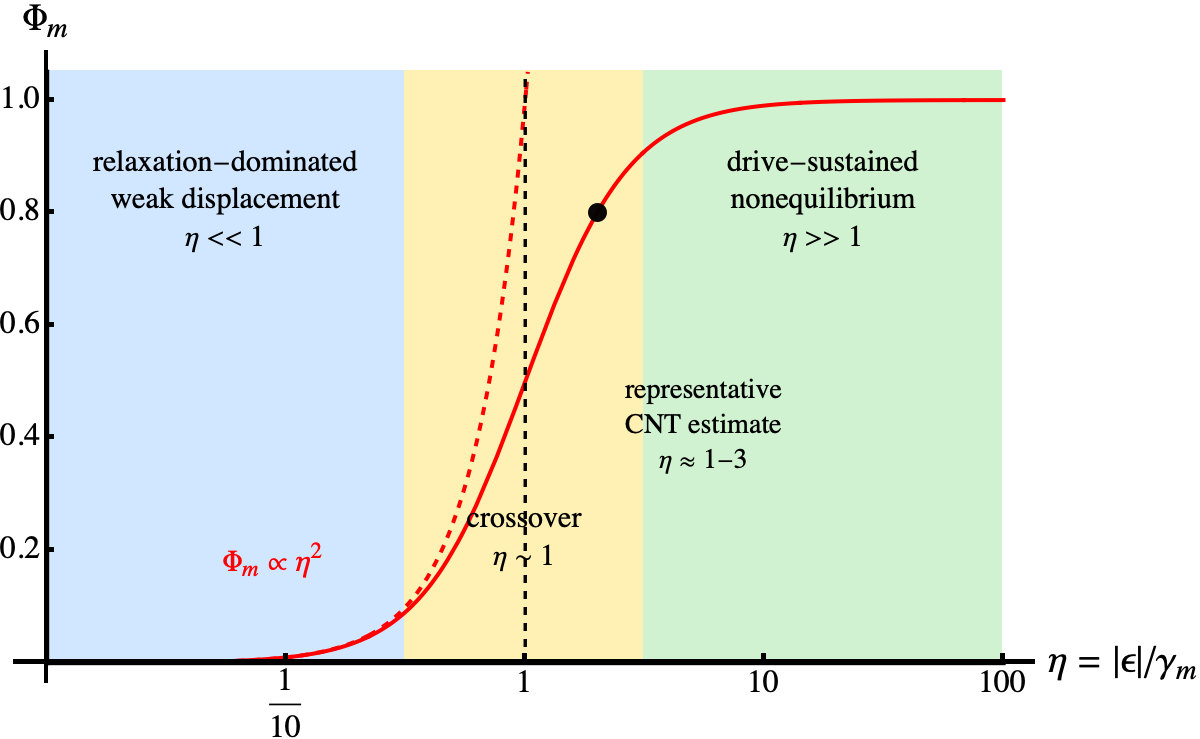}
    \caption{ Schematic illustration of the drive-induced crossover in the mechanical entropy flux.
The normalized mechanical entropy flux contribution is plotted as a function of the dimensionless drive strength
\(\eta=|\varepsilon|/\gamma_m\). The~solid red curve shows a schematic crossover form,
while the dashed red curve indicates the low-drive linear-response scaling
\(\Phi_m\propto \eta^2\). For~\(\eta\ll 1\), the~system is weakly displaced and the entropy budget is relaxation dominated. Near~\(\eta\sim 1\), the~system enters a crossover regime. For~\(\eta\gg 1\), the~resonator is driven far from its thermal reference state and the mechanical entropy flux becomes the dominant drive-sustained contribution. The~marked point indicates the representative CNT estimate \(\eta\simeq 1\)--\(3\) discussed in the text. \label{fig7}}
\end{figure}

A simple scaling analysis clarifies the competition between coherent hybridization and dissipation. Let the cooperativity-like ratio be defined as
\begin{equation} 
C=\frac{4g_1^2}{\gamma_m\Gamma_2}. 
\end{equation} 
For $C\ll 1$, the~spin and oscillator mainly relax through their own baths, and~entropy production is approximately additive. For~$C\gtrsim 1$, coherent exchange competes strongly with dissipation, spin--oscillator correlations become significant, and~the additive approximation fails. In~this crossover regime one expects the correlation contribution $\Pi_{\rm corr}$ to become important, especially under continuous coherent driving. Introducing the dimensionless drive parameter
\begin{equation} 
\eta=\frac{|\varepsilon|}{\gamma_m}, 
\end{equation} 
which provides a convenient measure of drive strength $\varepsilon= F_0x_{\mathrm{zpf}}/\hbar$ relative to mechanical damping, one distinguishes weakly displaced states ($\eta\ll 1$) from strongly nonequilibrium mechanical states ($\eta\gg 1$). The~entropy budget then crosses over from relaxation-dominated to drive-sustained, with~the mechanical entropy flux scaling as $\eta^2$ in the linear-response regime and remaining the dominant contribution once the resonator is held far from its thermal reference state (Figure~\ref{fig7}). We provide representative values for the parameters, which indicate that this thermal crossover should be experimentally accessible in suspended CNT devices. Using the length-dependent estimates of Ref.~\cite{chang_quantum_2025} for $Q=10^4$, one finds $\omega_m/2\pi \simeq 5.37~\mathrm{GHz}$, $221~\mathrm{MHz}$, and~$54~\mathrm{MHz}$ for $L=100$, $500$, and~$1000~\mathrm{nm}$, with~corresponding relaxation times $T_1 \simeq 0.29$, $7.4$, and~$29.6~\mu\mathrm{s}$ and, in~the relaxation-limited regime, $T_2 \simeq 2T_1$. At~$T=10~\mathrm{mK}$ these frequencies correspond to thermal occupations $\bar n_m \simeq 6\times10^{-12}$, $0.5$, and~$3.4$, respectively, showing explicitly the crossover from a near-ground-state mechanical mode in short CNTs to a thermally occupied mode in longer devices~\cite{chang_quantum_2025}. 

A complementary benchmark is provided by the suspended-CNT spin--phonon estimates in Ref.~\cite{palyi_spinorbitinduced_2012} who obtained $g_1/2\pi \simeq 0.56~\mathrm{MHz}$ for the spin qubit and $g_1/2\pi \simeq 0.49~\mathrm{MHz}$ for the transverse Kramers ($K_x$) qubit at $\omega_m/2\pi \simeq 500~\mathrm{MHz}$, together with a mechanical damping rate $\gamma_m \simeq 5\times10^4~\mathrm{s}^{-1}$ for $Q\simeq6.3\times10^4$. Combining these values with a representative transverse decoherence rate $\Gamma_2 \sim 10^5~\mathrm{s}^{-1}$ yields $C\sim10^4$, i.e.,~well within the correlation-dominated regime. In~the same work, the~strong-drive response was analyzed at $T=50~\mathrm{mK}$ with drive scale $\omega/2\pi\simeq0.027~\mathrm{MHz}$. Identifying $|\varepsilon|$ with this order of magnitude gives $\eta \simeq$ {1--3}%MDPI: We revised the minus sign into an endash to indicate the range, please confirm.
, which is consistent with a driven regime in which the mechanical entropy flux can dominate the steady-state entropy~budget. 

\section{Discussion} 
\label{sec:discussion} 
The framework developed above brings together three ingredients that are often treated separately: molecular spin qubits based on paramagnetic endohedral fullerenes, coherently driven suspended carbon nanotube mechanics, and~entropy production in open quantum systems. Their combination is compelling not only from the perspective of nonequilibrium thermodynamics but~also in relation to decoherence studies, quantum information flow, and~the control of hybrid quantum devices. First, PEF-filled CNTs provide a natural realization of a structured environment in which the distinction between ``system'' and ``bath'' is partially hierarchical rather than sharply binary. The~selected fullerene spin is the most localized degree of freedom, and~the addressed CNT flexural mode is an intermediate mesoscopic subsystem that can store, transfer, and~dissipate excitations. Moreover, the~remaining phononic and electromagnetic reservoirs form the broader environment. Entropy production in such a setting is richer than in a featureless Markov bath because coherent exchange, thermally activated transitions, and~reservoir-induced diffusion all operate on comparable footing. This makes the platform especially relevant for studying how irreversibility emerges when a localized qubit interacts with an environment that is itself structured, controllable, and~partially quantum. Second, the~Wigner-function formulation is particularly well suited to this problem because the mechanical degree of freedom is both experimentally measurable and theoretically tractable. In~realistic experiments one may not reconstruct the full joint density matrix of the spin--resonator system, but~one can often access mechanical quadratures, sideband spectra, linewidths, and~driven phase-space motion~\cite{weinbub_recent_2018, case_wigner_2008, bertet_direct_2002, leibfried_experimental_1996, lougovski_fresnel_2003, karlovets_possibility_2017, chang_quantum_2025}. These observables constrain the covariance matrix $\Sigma(t)$ and therefore the mechanical contribution to entropy generation. More broadly, the~phase-space picture provides a useful bridge between quantum control and open-system thermodynamics: it makes it possible to visualize the transition from coherent state preparation to decoherence-induced broadening and, in~the strong-drive regime, from~reversible spin--oscillator entanglement to genuine dissipative irreversibility. Thirdly, the~present framework is relevant to ongoing efforts in quantum computing and quantum information science. Hybrid spin--mechanical systems are of interest as candidate interfaces for quantum state transfer, transduction, and~quantum memory, and~their usefulness depends critically on understanding how coherence is degraded by vibrational and environmental channels~\cite{harneit_fullerenes_2017, rips_hartmann_prl_2013, pinto_readout_2020}. In~this context, entropy production provides more than a thermodynamic bookkeeping device: it offers a quantitative measure of how information about the spin qubit is redistributed into mechanical motion and ultimately lost to external reservoirs. The~same formalism can therefore be viewed as a tool for analyzing decoherence pathways, identifying operating regimes in which coherent hybridization dominates over dissipative loss, and~clarifying when a structured vibrational mode acts as a useful quantum resource rather than merely as a source of noise. From~this perspective, CNT--PEF devices are relevant not only as model systems for irreversibility but~also as testbeds for error mechanisms and control strategies in nanoscale quantum~architectures. 

The rotating-wave approximation is valid when the spin--phonon coupling is small compared with the fast frequencies that multiply the counter-rotating terms. In~the interaction picture, as~shown in Appendix~\ref{appA}, the~resonant terms \(a\sigma_+\) and \(a^\dagger\sigma_-\) oscillate at the detuning \(\Delta=\omega_s-\omega_m\), whereas the counter-rotating terms \(a\sigma_-\) and \(a^\dagger\sigma_+\) oscillate at \(\omega_s+\omega_m\). Thus, the~Jaynes--Cummings description requires \(g_1 \ll \omega_s+\omega_m\), with~the near resonant exchange regime further requiring \(|\Delta|=|\omega_s-\omega_m|\ll \omega_s,\omega_m\). The~leading effect of the discarded counter-rotating terms is a Bloch--Siegert type shift of the resonance condition, of~order \(\delta_{\rm BS}\sim g_1^2/(\omega_s+\omega_m)\), together with small virtual excitation corrections to the spin--mechanical eigenstates~\cite{breuer_theory_2009}. Efficient coherent exchange additionally requires \(|\Delta|\) to be comparable to or smaller than the relevant coupling and linewidth scales~\cite{breuer_theory_2009, chang_quantum_2025}. These corrections become non-negligible only as the system approaches the ultrastrong-coupling regime, \(g_1/\omega_m\gtrsim 0.05\)--\(0.1\), or~when the detuning is large enough that the near-resonant exchange picture no longer applies. In~that regime, one should replace the Jaynes--Cummings Hamiltonian by the full quantum Rabi Hamiltonian, retaining the complete interaction \(\hbar g_1(a+a^\dagger)\sigma_x\) rather than only its excitation-conserving part. The~entropy balance framework used here would remain applicable, but~the quantitative entropy production rates would be modified by Bloch--Siegert shifts, virtual spin--phonon excitations, and~additional correlation-mediated entropy channels. For~the representative CNT parameters discussed below, however, \(g_1/2\pi\sim 0.5\,{\rm MHz}\) and \(\omega_m/2\pi\sim 500\,{\rm MHz}\), so \(g_1/\omega_m\sim 10^{-3}\), and~the Bloch--Siegert correction is expected to be negligible on the scale of the dissipative rates considered~here.

The fullerene setting also provides a natural route to extensions beyond the effective two-level approximation adopted here. Both N@C$_{60}$ and P@C$_{60}$ possess additional spin sublevels and hyperfine structure that could support genuinely multilevel thermodynamic effects, including entropy redistribution among internal states, mode-selective relaxation pathways, and~possibly autonomous refrigeration or heat engine-like cycles under suitably engineered driving protocols. The~qubit treatment developed here should therefore be regarded as the lowest-order member of a broader class of spin--vibrational entropy problems in molecular nanomechanics. At~the same time, several aspects of the present treatment merit further refinement. The~Lindblad description assumes weak coupling to broad baths and therefore does not capture possible non-Markovian memory effects inherited from the nanotube environment, from~slow magnetic fluctuations, or~from additional structured vibrational modes. Likewise, the~exact Gaussian propagator for the oscillator is most directly applicable for linear damping and quadratic mechanical Hamiltonians. Strong mechanical nonlinearities, intermode coupling, transport-induced backaction, or~drive-induced anharmonicities would require extensions beyond the present phase space treatment. In~addition, the~microscopic derivation of the spin--phonon coupling for a specific fullerene species in a specific nanotube geometry remains an open materials-level problem. The~description used here is therefore best viewed as an effective, physically grounded framework guided by the established physics of suspended CNT spin devices and fullerene peapods, rather than as a complete microscopic model of any one experimental~realization. 

These considerations also suggest several natural directions for future work. A~first step is to incorporate the full hyperfine-resolved structure of N@C$_{60}$ or P@C$_{60}$, thereby replacing the effective qubit with a multilevel spin manifold. A~second is to include multiple CNT vibrational modes and examine mode-selective entropy currents, intermode correlations, and~their influence on decoherence. A~third is to move beyond Markovian dissipation and analyze memory effects in the entropy balance, especially in regimes where the mechanical mode mediates long-lived backaction on the spin sector. A~fourth is to connect the effective parameters of the theory more directly to experimentally realistic geometries by modeling, for~example, magnetic-gradient control from a nearby nanomagnet or a magnetized AFM tip and by incorporating device-specific estimates of damping, dephasing, and~thermal occupation. More generally, it would be valuable to relate entropy-production diagnostics to standard figures of merit in quantum information processing, such as state-transfer fidelity, coherence time, and~gate error, thereby linking nonequilibrium thermodynamics more directly to the performance of hybrid molecular quantum devices. Taken together, these results suggest that suspended CNTs filled with paramagnetic endohedral fullerenes offer an unusually versatile setting in which to study the interplay of coherent control, decoherence, and~irreversibility. They provide a platform where entropy flow can be analyzed not only as a thermodynamic quantity but~also as a diagnostic of information loss and environment-induced degradation in a hybrid quantum system. In~that sense, the~present framework is relevant both to the foundations of nonequilibrium quantum thermodynamics and to the practical problem of understanding how structured vibrational environments shape the operation of nanoscale~qubits.

\section{Conclusions}
\label{sec:Conclusions}
We have presented a theoretical framework for entropy production in a hybrid quantum nanostructure composed of paramagnetic endohedral fullerene qubits encapsulated in a suspended carbon nanotube resonator. The~approach combines an effective spin--phonon Hamiltonian motivated by suspended CNT spin-control schemes with a Wigner-function treatment of the driven, damped mechanical mode and an entropy-balance analysis based on Lindblad quantum~dynamics.

The main result is a transparent description of how irreversibility is distributed among coherent spin--vibrational exchange, mechanical diffusion, spin relaxation, and~dephasing. The~Wigner representation isolates the respective roles of drive, damping, diffusion, and~spin-conditioned phase space splitting in the CNT resonator, while the Lindblad entropy balance identifies the corresponding entropy flux and non-negative entropy production of the full hybrid system. This makes it possible to distinguish oscillator-dominated and spin-dominated entropy production regimes, to~analyze thermal and drive-induced crossovers between them, and~to identify the parameter ranges in which spin--oscillator correlations make an appreciable contribution to the entropy~budget.

A further outcome of the present analysis is that it connects entropy production directly to the problem of decoherence in a structured hybrid environment. In~the CNT--PEF platform, information initially stored in the localized spin qubit can be transferred coherently to the vibrational mode and then dissipated into external reservoirs. Entropy production therefore provides a quantitative way to track how coherence is degraded, how spin--oscillator correlations are generated and lost, and~how a structured vibrational mode transitions from a useful quantum intermediary to a channel of irreversible information loss. More generally, the~results show how phase-space methods and open-system thermodynamics can be fruitfully combined in hybrid devices where localized spins interact with controllable vibrational environments. Suspended fullerene-filled nanotubes are therefore of interest not only as experimentally relevant platforms for studying irreversibility at the single-molecule level but~also as candidate architectures for quantum information processing, coherent state transfer, and~hybrid quantum control. In~this broader setting, entropy-production diagnostics may complement more standard quantum-information figures of merit by helping identify operating regimes in which coherent hybridization remains dominant and dissipative losses remain~controllable.

Overall, these results demonstrate that PEF-filled suspended CNTs offer a versatile setting in which to study the interplay of coherent control, decoherence, and~irreversibility in a structured quantum environment. They provide a platform where entropy flow can be interpreted simultaneously as a thermodynamic quantity and as a diagnostic of information redistribution and loss in a hybrid quantum~device.

\vspace{0.5\baselineskip}

\begin{acknowledgments}
The author acknowledge financial support for this work from Tufts University Faculty Research Award.  
\end{acknowledgments}

\newpage

\bibliography{NewCitation}% Produces the bibliography via BibTeX.

\clearpage
\onecolumngrid

\begin{center}
{\bf\large{Appendices for Entropy Production from Spin–Vibrational Coupling in Endohedral-Fullerene Qubits Encapsulated in Suspended Carbon Nanotubes}}
\end{center}

\appendix

\section{Derivation of the Effective Jaynes--Cummings Hamiltonian}
\label{appA}

\setcounter{equation}{0}
\renewcommand{\theequation}{A\arabic{equation}}

In this appendix we derive the effective Jaynes--Cummings (JC) Hamiltonian used in the main text for a paramagnetic endohedral fullerene (PEF) spin qubit coupled to a flexural mode of a suspended carbon nanotube (CNT). The~derivation follows the logic discussed in the main text: we begin from the laboratory-frame Hamiltonian, transform to a suitable rotating frame, and~then apply the rotating-wave approximation (RWA). Since the goal here is to obtain the coherent spin--phonon Hamiltonian, we do not include the environmental bath, which is treated in the main text using the Lindblad~formalism.

\medskip
\noindent\textit{{Laboratory-frame Hamiltonian}%MDPI: 1. Please confirm if the italics are necessary; if not, please remove them. 2. Please confirm if indentation should be added; of if the italics contents should be formatted as a list or a subsection. The following highlights are the same. Please check and confirm all.
}. The~coherent PEF--CNT dynamics is described by
\begin{equation}
H_{\rm lab}(t)=H_0+H_{\rm int}+H_d^{\rm lab}(t),
\label{A1}
\end{equation}
with
\begin{equation}
H_0=\frac{\hbar\omega_s}{2}\sigma_z+\hbar\omega_m a^\dagger a,
\label{A2}
\end{equation}
and
\begin{equation}
H_{\rm int}=\hbar g_1(a+a^\dagger)\sigma_x, \quad H_d^{\rm lab}(t)
=-F_0x_{\rm zpf}(a+a^\dagger)\cos(\omega_d t)
\label{A3}
\end{equation}
Here $\omega_s$ is the PEF spin splitting, $\omega_m$ is the CNT mechanical frequency, $a$ and $a^\dagger$ are the annihilation and creation operators of the flexural mode, and~$\sigma_x,\sigma_z$ are Pauli operators acting in the effective two-level spin subspace. Writing
\vspace{-3pt}\begin{equation}
\sigma_x=\sigma_+ + \sigma_-,
\label{A4}
\end{equation}
the interaction term becomes
\vspace{-3pt}\begin{align}
H_{\rm int}
&=\hbar g_1(a+a^\dagger)(\sigma_+ + \sigma_-)
\nonumber\\[3pt]
&=\hbar g_1\left(a\sigma_+ + a\sigma_- + a^\dagger\sigma_+ + a^\dagger\sigma_-\right).
\label{A5}
\end{align}

\medskip
\noindent\textit{{Rotating-frame Hamiltonian}}. To~isolate the slowly varying near-resonant terms, we move to a frame rotating at the mechanical frequency $\omega_m$. We introduce the unitary transformation
\begin{equation}
U(t)=\exp\!\left[i\omega_m t\left(a^\dagger a+\frac{\sigma_z}{2}\right)\right],
\label{A6}
\end{equation}
and define the transformed Hamiltonian  as~\cite{schlosshauer_decoherence_2007, breuer_theory_2009}
\vspace{-3pt}\begin{equation}
\tilde H(t)=U(t)H_{\rm lab}(t)U^\dagger(t)-i\hbar\,U(t)\frac{d}{dt}U^\dagger(t).
\label{A7}
\end{equation}
The relevant operators transform according to
\begin{align}
U a U^\dagger &= a\,e^{-i\omega_m t},
\label{A8}\\
U a^\dagger U^\dagger &= a^\dagger e^{i\omega_m t},
\label{A9}\\
U \sigma_+ U^\dagger &= \sigma_+ e^{i\omega_m t},
\label{A10}\\
U \sigma_- U^\dagger &= \sigma_- e^{-i\omega_m t}.
\label{A11}
\end{align}
Applying Equation~\eqref{A6} to the free Hamiltonian gives
\begin{equation}
U H_0 U^\dagger - i\hbar\,U\frac{d}{dt}U^\dagger
= \frac{\hbar(\omega_s-\omega_m)}{2}\sigma_z
\equiv \frac{\hbar\Delta}{2}\sigma_z,
\label{A12}
\end{equation}
where
\begin{equation}
\Delta\equiv \omega_s-\omega_m
\label{A13}
\end{equation}
is the detuning~frequency.

The interaction term transforms as
\begin{align}
U H_{\rm int} U^\dagger
&=\hbar g_1
\Bigl[
a\sigma_+ + a^\dagger\sigma_-
+a\sigma_-e^{-2i\omega_m t}
+a^\dagger\sigma_+e^{2i\omega_m t}
\Bigr].
\label{A14}
\end{align}
The first two terms are time independent in this frame and correspond to the resonant exchange of a spin excitation and a phonon. The~last two terms oscillate rapidly at frequency $2\omega_m$ and are the counter-rotating contributions. Similarly, the~drive term transforms as
\begin{align}
U H_d^{\rm lab}(t) U^\dagger
&= -F_0x_{\rm zpf}
\left(ae^{-i\omega_m t}+a^\dagger e^{i\omega_m t}\right)
\cos(\omega_d t) \nonumber\\
&= -\frac{F_0x_{\rm zpf}}{2}
\Big[
a e^{-i(\omega_m-\omega_d)t}
+a^\dagger e^{i(\omega_m-\omega_d)t}
\nonumber\\
&\hspace{2.5cm}
+a e^{-i(\omega_m+\omega_d)t}
+a^\dagger e^{i(\omega_m+\omega_d)t}
\Big].
\end{align}
For a near-resonant drive, \(\omega_d\simeq\omega_m\), the~terms oscillating at
\(\omega_m+\omega_d\) are rapidly varying and are neglected under the RWA. Thus
\vspace{-3pt}\begin{equation}
\tilde H_d^{\rm RWA}(t)
= -\frac{F_0x_{\rm zpf}}{2}
\left[
a e^{-i\delta_d t}
+a^\dagger e^{i\delta_d t}
\right],
\label{A15}
\end{equation}
where \(\delta_d=\omega_m-\omega_d\). On~exact resonance, \(\delta_d=0\), this reduces to
\begin{equation}
\tilde H_d^{\rm RWA}
= -\frac{F_0x_{\rm zpf}}{2}(a+a^\dagger).
\end{equation}

\medskip
\noindent\textit{{System Hamiltonian in the rotating-wave approximation}}. When the system is close to resonance and the coupling is weak, $|\Delta|\ll \omega_m$, $g_1\ll \omega_m,\omega_s$, the~rapidly oscillating terms average to zero on the timescale of the slow dynamics. Under~the RWA we therefore keep only the resonant terms and obtain the Jaynes--Cummings Hamiltonian
\begin{equation}
H_{\rm JC}(t)=
\frac{\hbar\Delta}{2}\sigma_z
+\hbar g_1\left(a\sigma_+ + a^\dagger\sigma_-\right)
-\frac{F_0x_{\rm zpf}}{2}(a+a^\dagger).
\end{equation}

Thus, the~coherent spin--phonon dynamics reduces to the standard Jaynes--Cummings interaction. This is the form used throughout the main text. It shows that, near~resonance, the~dominant coherent process is the exchange of a single phonon in the CNT mode with a single spin excitation of the PEF qubit, with~coupling strength $g_1$, while the detuning $\Delta$ controls the deviation from exact~resonance.

\section{Gaussian Propagator and Covariance Dynamics}
\label{appB}

\setcounter{equation}{0}
\renewcommand{\theequation}{B\arabic{equation}}
In this appendix, we derive the explicit Gaussian kernel entering Equation~\eqref{eq:Wprop}, together with the corresponding covariance matrix and mean-trajectory~dynamics.

For a linear quantum Brownian oscillator, the~Wigner function remains Gaussian if the initial state is Gaussian. Let the drift matrix in Equation~\eqref{eq:Sigmaeq} be
\begin{equation}
A = 
\begin{pmatrix}
0 & 1/m \\
-m\omega_m^2 & -\gamma_m
\end{pmatrix},
\end{equation}
and let $D$ denote the diffusion matrix. Then the covariance matrix evolves as
\begin{equation}
\Sigma(t) = e^{At}\Sigma(0)e^{A^T t} + \int_0^t ds\; e^{A(t-s)} D e^{A^T(t-s)}.
\end{equation}
The mean phase-space coordinate obeys
\begin{equation}
\bar{\mathbf{x}}(t) = e^{At}\bar{\mathbf{x}}(0) + \int_0^t ds\; e^{A(t-s)}\mathbf{f}(s),
\end{equation}
with $\mathbf{f}(s)=(0,F_d(s))^T$. Inserting these expressions into Equation~\eqref{eq:KWgaussian} yields the exact Gaussian propagator used in the main~text.

\section{Wigner Function of the Reduced Mechanical State in the Weak-Drive Regime}
\label{appC}

\setcounter{equation}{0}
\renewcommand{\theequation}{C\arabic{equation}}

In this appendix we derive the reduced mechanical Wigner function used in the main text for the resonant weak-drive regime. Starting from the pure Jaynes--Cummings state
\begin{equation}
|\psi(t)\rangle=\cos(gt)\,|e,0\rangle-i\sin(gt)\,|g,1\rangle,
\label{C1}
\end{equation}
the reduced mechanical density operator obtained by tracing over the spin is
\begin{equation}
\rho_m^{(0)}(t)=\mathrm{Tr}_s\!\left(|\psi(t)\rangle\langle\psi(t)|\right)
=\cos^2(gt)|0\rangle\langle 0|+\sin^2(gt)|1\rangle\langle 1|.
\label{C2}
\end{equation}
Because the Wigner transform is linear in the density operator, the~corresponding reduced mechanical Wigner function is
\begin{equation}
W_m(q,p,t)=\cos^2(gt)\,W_0(q,p)+\sin^2(gt)\,W_1(q,p),
\label{C3}
\end{equation}
where $W_0$ and $W_1$ are the Wigner functions of the harmonic-oscillator ground state and first excited state, respectively.

\subsection*{C1: Definition of the Wigner Function}
For a mechanical density operator $\rho_m$, the~Wigner function is defined as
\begin{equation}
W(q,p)=\frac{1}{2\pi\hbar}\int_{-\infty}^{\infty} d\xi\;
e^{-ip\xi/\hbar}
\left\langle q+\frac{\xi}{2}\right|\rho_m\left|q-\frac{\xi}{2}\right\rangle.
\label{C4}
\end{equation}
Applying this definition to Equation~\eqref{C2} immediately gives Equation~\eqref{C3} since
\vspace{-3pt}\begin{align}
W_m(q,p,t)
&=\frac{1}{2\pi\hbar}\int_{-\infty}^{\infty} d\xi\;
e^{-ip\xi/\hbar}
\left\langle q+\frac{\xi}{2}\right|\rho_m^{(0)}(t)\left|q-\frac{\xi}{2}\right\rangle
\nonumber\\[3pt]
&=\cos^2(gt)\,
\frac{1}{2\pi\hbar}\int_{-\infty}^{\infty} d\xi\;
e^{-ip\xi/\hbar}
\psi_0\!\left(q+\frac{\xi}{2}\right)\psi_0^*\!\left(q-\frac{\xi}{2}\right)
\nonumber\\
&\quad+
\sin^2(gt)\,
\frac{1}{2\pi\hbar}\int_{-\infty}^{\infty} d\xi\;
e^{-ip\xi/\hbar}
\psi_1\!\left(q+\frac{\xi}{2}\right)\psi_1^*\!\left(q-\frac{\xi}{2}\right),
\label{C5}
\end{align}
where $\psi_0(q)=\langle q|0\rangle$ and $\psi_1(q)=\langle q|1\rangle$ are the harmonic-oscillator eigenfunctions. The~two integrals in Equation~\eqref{C5} are precisely $W_0$ and $W_1$.

\subsection*{C2: Harmonic-Oscillator Wavefunctions}
Let
\begin{equation}
\ell=\sqrt{\frac{\hbar}{m\omega_m}}
\label{C6}
\end{equation}
denote the oscillator length of the CNT flexural mode, and~$m$ its effective mass. The~normalized ground-state and first-excited-state wavefunctions are
\begin{align}
\psi_0(q)
&=\frac{1}{(\pi \ell^2)^{1/4}}
\exp\!\left(-\frac{q^2}{2\ell^2}\right),
\label{C7}\\[4pt]
\psi_1(q)
&=
\frac{\sqrt{2}\,q}{\ell(\pi \ell^2)^{1/4}}
\exp\!\left(-\frac{q^2}{2\ell^2}\right).
\label{C8}
\end{align}

\subsection*{C3: Ground-State Wigner Function $W_0$}
Substituting $\rho_m=|0\rangle\langle 0|$ into Equation~\eqref{C4} gives
\begin{equation}
W_0(q,p)=\frac{1}{2\pi\hbar}\int_{-\infty}^{\infty} d\xi\;
e^{-ip\xi/\hbar}
\psi_0\!\left(q+\frac{\xi}{2}\right)\psi_0^*\!\left(q-\frac{\xi}{2}\right).
\label{C9}
\end{equation}
Using Equation~\eqref{C7},
\begin{align}
\psi_0\!\left(q+\frac{\xi}{2}\right)\psi_0^*\!\left(q-\frac{\xi}{2}\right)
&=\frac{1}{\sqrt{\pi}\,\ell}
\exp\!\left[-\frac{\left(q+\xi/2\right)^2+\left(q-\xi/2\right)^2}{2\ell^2}\right]
\nonumber\\
&=\frac{1}{\sqrt{\pi}\,\ell}
\exp\!\left(-\frac{q^2}{\ell^2}-\frac{\xi^2}{4\ell^2}\right).
\label{C10}
\end{align}
Therefore
\begin{equation}
W_0(q,p)=\frac{e^{-q^2/\ell^2}}{2\pi\hbar\,\sqrt{\pi}\,\ell}
\int_{-\infty}^{\infty} d\xi\;
\exp\!\left(-\frac{\xi^2}{4\ell^2}-\frac{i p\xi}{\hbar}\right).
\label{C11}
\end{equation}
Evaluating the Gaussian integral yields
\begin{equation}
W_0(q,p)=\frac{1}{\pi\hbar}
\exp\!\left(-\frac{q^2}{\ell^2}-\frac{\ell^2 p^2}{\hbar^2}\right).
\label{C12}
\end{equation}

\subsection*{C4: First-Excited-State Wigner Function $W_1$}
For the one-phonon state,
\begin{equation}
W_1(q,p)=\frac{1}{2\pi\hbar}\int_{-\infty}^{\infty} d\xi\;
e^{-ip\xi/\hbar}
\psi_1\!\left(q+\frac{\xi}{2}\right)\psi_1^*\!\left(q-\frac{\xi}{2}\right).
\label{C13}
\end{equation}
Using Equation~\eqref{C8},
\begin{align}
\psi_1\!\left(q+\frac{\xi}{2}\right)\psi_1^*\!\left(q-\frac{\xi}{2}\right)
&=
\frac{2}{\ell^2}
\left(q+\frac{\xi}{2}\right)\left(q-\frac{\xi}{2}\right)
\psi_0\!\left(q+\frac{\xi}{2}\right)\psi_0^*\!\left(q-\frac{\xi}{2}\right)
\nonumber\\
&=
\frac{2}{\ell^2}
\left(q^2-\frac{\xi^2}{4}\right)
\frac{1}{\sqrt{\pi}\,\ell}
\exp\!\left(-\frac{q^2}{\ell^2}-\frac{\xi^2}{4\ell^2}\right).
\label{C14}
\end{align}
Substituting into Equation~\eqref{C13} gives
\begin{equation}
W_1(q,p)=
\frac{2\,e^{-q^2/\ell^2}}{2\pi\hbar\,\ell^2\sqrt{\pi}\,\ell}
\int_{-\infty}^{\infty} d\xi\;
\left(q^2-\frac{\xi^2}{4}\right)
\exp\!\left(-\frac{\xi^2}{4\ell^2}-\frac{i p\xi}{\hbar}\right).
\label{C15}
\end{equation}
Carrying out the Gaussian integrals yields
\begin{equation}
W_1(q,p)=\frac{1}{\pi\hbar}
\left[
2\left(\frac{q^2}{\ell^2}+\frac{\ell^2 p^2}{\hbar^2}\right)-1
\right]
\exp\!\left(-\frac{q^2}{\ell^2}-\frac{\ell^2 p^2}{\hbar^2}\right).
\label{C16}
\end{equation}

\subsection*{C5: Final Form of the Reduced Mechanical Wigner Function}
Substituting Equations~\eqref{C12} and \eqref{C16} into Equation~\eqref{C3} gives
\begin{align}
W_m(q,p,t)
&=
\cos^2(gt)\,
\frac{1}{\pi\hbar}
\exp\!\left(-\frac{q^2}{\ell^2}-\frac{\ell^2 p^2}{\hbar^2}\right)
\nonumber\\
&\quad+
\sin^2(gt)\,
\frac{1}{\pi\hbar}
\left[
2\left(\frac{q^2}{\ell^2}+\frac{\ell^2 p^2}{\hbar^2}\right)-1
\right]
\exp\!\left(-\frac{q^2}{\ell^2}-\frac{\ell^2 p^2}{\hbar^2}\right).
\label{C17}
\end{align}
Equivalently,
\begin{equation}
W_m(q,p,t)=\cos^2(gt)\,W_0(q,p)+\sin^2(gt)\,W_1(q,p),
\label{C18}
\end{equation}
which is the expression quoted in the main~text.

It is often convenient to introduce dimensionless phase-space variables
\begin{equation}
Q=\frac{q}{\ell},\qquad P=\frac{\ell p}{\hbar},
\label{C19}
\end{equation}
in terms of which
\begin{align}
W_0(Q,P)
&=\frac{1}{\pi\hbar}e^{-(Q^2+P^2)},
\label{C20}\\
W_1(Q,P)
&=\frac{1}{\pi\hbar}\Bigl[2(Q^2+P^2)-1\Bigr]e^{-(Q^2+P^2)}.
\label{C21}
\end{align}
These expressions make clear that the reduced mechanical phase-space distribution oscillates between the Gaussian vacuum profile and the nonclassical one-phonon profile as the excitation is coherently exchanged between the spin and the CNT vibrational~mode.

\end{document}